\documentclass[amsmath,amssymb,prb,preprintnumbers,superscriptaddress,twocolumn,showpacs,longbibliography]{revtex4-1}

\usepackage{graphicx}

\usepackage{bm}         
\usepackage{amssymb}
\usepackage{amsmath}
\usepackage{epsfig}
\usepackage[usenames]{color}

\newcommand{\gapprox}{{\scriptscriptstyle\stackrel{>}{\sim}\,}}
\newcommand{\lapprox}{{\scriptscriptstyle\stackrel{<}{\sim}\,}}

\newcommand{\hastfill}{\hspace*{\fill}}

\newif\ifcom
\newif\ifdel

\comtrue               %
\deltrue                   %

\begin{document}

%


\title{NanoSQUIDs: Basics \& recent advances}

\author{M.~J.~Mart\'{i}nez-P\'{e}rez}
\author{D.~Koelle}
\affiliation{%
Physikalisches Institut -- Experimentalphysik II and Center for Quantum Science in LISA$^+$,
Universit\"at T\"ubingen,
Auf der Morgenstelle 14,
D-72076 T\"ubingen, Germany}

\date{\today}

\begin{abstract}
Superconducting Quantum Interference Devices (SQUIDs) are one of the most popular  devices in superconducting electronics.
They  combine the Josephson effect with the quantization of magnetic flux in superconductors.
This gives rise to one of the most beautiful manifestations of macroscopic quantum coherence in the solid state.
In addition, SQUIDs are extremely sensitive sensors allowing to transduce magnetic flux into measurable electric signals.
As a consequence, any physical observable that can be converted into magnetic flux, e.g., current, magnetization, magnetic field or position, becomes easily accessible to SQUID sensors.
In the late 1980's it became clear that downsizing the dimensions of SQUIDs to the nanometric scale would encompass an enormous increase of their sensitivity to localized tiny magnetic signals.
Indeed, nanoSQUIDs opened the way to the investigation of, e.g., individual magnetic nanoparticles or surface magnetic states with unprecedented sensitivities.
The purpose of this review is to present a detailed survey of microscopic and nanoscopic SQUID sensors.
We will start by discussing the principle of operation of SQUIDs, placing the emphasis on their application as ultrasensitive detectors for small localized magnetic signals.
We will continue by reviewing a number of existing devices based on different kinds of Josephson junctions and materials, focusing on their advantages and drawbacks.
The last sections are left for applications of nanoSQUIDs in the fields of scanning SQUID microscopy and magnetic particle characterization, putting special stress on the investigation of individual magnetic nanoparticles.
\end{abstract}

\maketitle

\tableofcontents

\section{Introduction}
\label{sec:Introduction}

The superconducting quantum interference device (SQUID) consists of a superconducting ring intersected by one (rf SQUID) or two (dc SQUID) Josephson junctions.
%
SQUIDs constitute, still at present, the most sensitive sensors for magnetic flux in the solid state \cite{Clarke04,Kleiner04}. 
For more than 50 years, a plethora of devices exploiting this property have been envisioned, fabricated and used in many fields of applications \cite{Clarke06}.
These devices include voltmeters, current amplifiers, metrology standards, motion sensors and magnetometers.
One of the key applications of SQUIDs is in magnetometry.
Here, a superconducting input circuit (flux transformer) picks up the magnetic flux density $B$, captured by superconducting pick-up loops of some mm$^2$ or cm$^2$ area, and the induced current is then (typically inductively) coupled to a SQUID. 
The figure of merit of SQUID magnetometers is the field resolution $\sqrt{S_B}=\sqrt{S_\Phi}/A_{\rm eff}$, which can reach values down to about $1 \rm{fT}/\sqrt{\rm Hz}$.
Here, $S_\Phi$ is the spectral density of flux noise of the SQUID and $A_{\rm eff}$ is the effective area of the magnetometer.

To ensure good coupling from an input circuit to a SQUID, typically thin film multiturn input coils are integrated on top of a washer-type SQUID loop.
Typical thin film washer SQUIDs have lateral outer dimensions of several $100\,\mu$m, the inner hole size is several tens of $\mu$m and the lateral size of the Josephson junctions is several $\mu$m.
Such devices are fabricated by conventional thin film technology, including micropattering by photolithography.
With the development of a mature junction technology, based on sandwich-type Nb/Al-AlO$_x$/Nb junctions in the 1980s \cite{Gurvitch83}, Nb based dc SQUIDs became the most commonly used type of devices for various applications.
At the same time, first attempts were started to further miniaturize the lateral dimensions of SQUIDs, including the Josephson junctions \cite{Voss80}.
This was made possible by advances in nanolithography \cite{Broers76} and was motivated by the development of the theory for thermal noise in the dc SQUID \cite{Tesche77}, which showed that the energy resolution $\varepsilon =S_\Phi/(2L)$ of dc SQUIDs can be improved by reducing the SQUID loop inductance $L$ and junction capacitance $C$, to eventually reach and explore quantum limited resolution of such devices \cite{Koch81}.
These developments have triggered the realization of miniaturized dc SQUIDs for the investigating of small magnetic particles and for imaging of magnetic field distributions by scanning SQUID microscopy to combine high sensitivity to magnetic flux with high spatial resolution. 
In 1984, Ketchen {\it et al.} \cite{Ketchen84} presented the first SQUID microsusceptometer devoted to detect the tiny signal produced by micron-sized magnetic objects, and in 1983 Rogers and Bermon developed the first system to produce  2-dimensional scans of magnetic flux structures in superconductors \cite{Rogers83}.
Both developments were pushed further in the 1990s.
Wernsdorfer {\it et al.} \cite{Wernsdorfer95,Wernsdorfer01} used micron-sized SQUIDs to perform experiments on the magnetization reversal of nanometric particles, which were placed directly on top of the SQUIDs.
At the same time, scanning SQUID microscopes with miniaturized SQUIDs and/or pickup loop structures have been developed, at that time with focus on studies on the pairing symmetry in the high transition temperature (high-$T_c$) cuprate superconductors \cite{Kirtley10}.
Since then much effort has been dedicated to the further miniaturization of SQUID devices and to the optimization of their noise characteristics \cite{Granata16}.

Studies on the properties of small spin systems, such as magnetic nanoparticles (MNPs) and single molecule magnets (SMMs), have fueled the development of new magnetic sensors for single particle detection and imaging with improved performance.
Many of the recent advances in this field include the development of magneto-optical techniques based on nitrogen vacancy centers in diamond \cite{Schaefer-Nolte14,Thiel16} or the use of carbon nanotubes (CNTs) as spin detectors \cite{Ganzhorn13}.
Alternatively, miniature magnetometers, based on either microHall bars \cite{Lipert10} or micro- and nanoSQUIDs, provide direct measurement of the stray magnetic fields generated by the particle under study, making the interpretation of the results much more direct and simple.
While their sensitivity deteriorates rapidly when Hall sensors are reduced to the submicron size, miniaturized SQUID-based sensors can theoretically reach quantum limited resolution.

In this review, we give an overview on some basics of nanoSQUIDs
%
\footnote{The term nanoSQUID denotes strongly miniaturized thin film SQUIDs with lateral dimensions in the submicrometer range. However, some devices described here and also various statements made also refer to slightly larger structures, which sometimes are denoted as microSQUIDs. Throughout the text, we do not make this discrimination.} 
%
and recent advances in the field.
After a brief description of some SQUID basics in section \ref{sec:SQUIDbasics}, we will review in section \ref{sec:nanoSQUIDs} important design considerations for optimizing nanoSQUID performance 
and the state of the art in fabrication and performance of nanoSQUIDs based on low-$T_c$ 
and high-$T_c$ 
superconductors, with emphasis on the various types of Josephson junctions used.
Subsequently, we will review important applications of nanoSQUIDs, divided into two sections:
Section \ref{sec:MNP-detection} gives an overview on applications of nanoSQUIDs for magnetic particle detection, and section \ref{sec:SSM} addresses nanoSQUIDs for scanning SQUID microscopy.
We will conclude with a short section \ref{sec:Summary}, which gives a summary and outlook.

\section{SQUIDs: some basic considerations}
\label{sec:SQUIDbasics}

The working principle of a SQUID is based on two fundamental phenomena in superconductors, the fluxoid quantization and the Josephson effect.
The fluxoid quantization arises from the quantum nature of superconductivity, as the macroscopic wave function describing the whole ensemble of Cooper pairs shall not interfere destructively.
This leads to the quantization of the magnetic flux $\Phi$ threading a superconducting loop \cite{London50}, in units of the magnetic flux quantum $\Phi_0 = h/2e \approx 2.07 \times 10^{-15}\,$Vs.

The Josephson effect \cite{Josephson62,Anderson63} results from the overlap of the macroscopic wave functions between two superconducting electrodes at a weak link forming the Josephson junction (JJ).
The supercurrent $I_{\rm s}$ through the weak link and the voltage drop $U$ across it satisfy the Josephson relations 
%
\begin{equation}
I_{\rm s}(t) = I_0 \sin \delta(t)\quad({\rm a})\qquad
U(t) = \frac{\Phi_0}{2\pi}\; \dot{\delta}\quad({\rm b}) 
\quad  ,	
\label{eq:Jos}
\end{equation}
%
with the gauge-invariant phase difference $\delta$ between the macroscopic wave functions of both superconductors and the maximum attainable supercurrent $I_0$; the dot refers to the time derivative.
The simple sinusoidal current-phase relation (CPR), Eq.~(\ref{eq:Jos}(a)), is found for many kinds of JJs.
However, some JJ types exhibit a non-sinusoidal CPR, which can even be multivalued \cite{Likharev79}.

\subsection{Resistively and capacitively shunted junction model}
\label{subsec:RCSJ-model}

\begin{figure}[t]
%
\includegraphics[width=0.7\columnwidth]{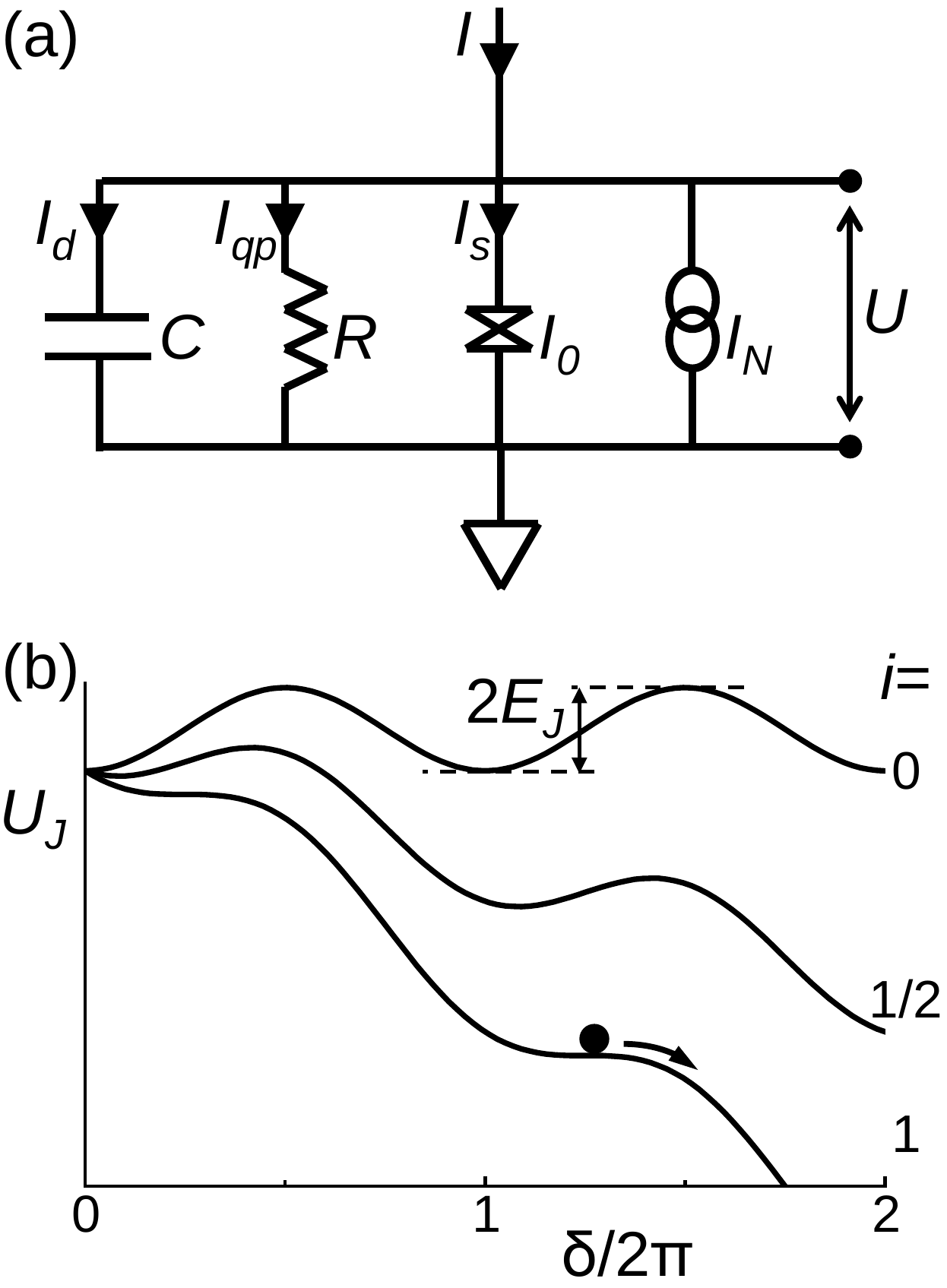}
%
\caption{RCSJ model: (a) Equivalent circuit.
(b) Tilted washboard potential for different normalized bias currents $i$.}
\label{Fig:1}
%
\end{figure}

A very useful approach to describe the phase dynamics of a JJ is the resistively and capacitively shunted junction (RCSJ) model \cite{Stewart68,McCumber68,Chesca-SHB-2}.
Within this model, the current flow is split into three parallel channels  [Fig.~\ref{Fig:1}(a)]: (i) a supercurrent $I_{\rm s}$ [Eq.~(\ref{eq:Jos}(a))], (ii) a dissipative quasiparticle current $I_{\rm qp} = U/R$ across an ohmic resistor $R$ and (iii) a displacement current $I_{\rm d} = C\;\partial U/\partial t$ across the junction capacitance $C$.
A finite temperature $T$ is included as a thermal current noise source $I_{\rm N}$ from the resistor.
With Kirchhoff's law and Eq.~(\ref{eq:Jos}(b)), one obtains the equation of motion for the phase difference $\delta$ 
%
\begin{equation}
I + I_{\rm N} = I_0 \sin\delta + \frac{\Phi_0}{2\pi R}\;\dot{\delta} + \frac{\Phi_0 C}{2\pi}\;\ddot{\delta} \quad  .	
\label{eq:DGL-delta}
\end{equation}
%
This is equivalent to the equation of motion of a point-like particle moving in a tilted washboard potential [Fig.~\ref{Fig:1}(b)]
%
\begin{equation}
U_{\rm J} = E_{\rm J} {(1-\cos\delta) - (i+i_{\rm N}) \delta}    \quad  ,	
\label{eq:twbp}
\end{equation}
%
with normalized currents $i=I/I_0$, $i_{\rm N}=I_{\rm N}/I_0$ and the Josephson coupling energy $E_{\rm J} = I_0\Phi_0/(2\pi)$.
In this analogy, the mass, friction coefficient, driving force (tilting the potential) and velocity correspond to $C$, $1/R$, $I$ and $U$, respectively.
Hysteresis in the current voltage characteristics (IVC), i.e. bias current $I$ vs time averaged voltage $V=\langle U\rangle$, can be understood as a consequence of the particle's inertia: the dissipative state $\langle\dot{\delta}\rangle \propto V \neq 0$ is achieved once the metastable minima of the washboard potential disappear at $I\ge I_0$.
If $I$ is decreased from $I > I_0$, the particle becomes retrapped at $I_{\rm r} < I_0$, leading to a hysteretic IVC.
This behavior can be quantified by the Stewart-McCumber parameter 
%
\begin{equation}
\beta_C \equiv \frac{2\pi}{\Phi_0}\; I_0 R^2 C      \quad  .	
\label{eq:betaC}
\end{equation}
%
In order to obtain a non-hysteretic IVC, $\beta_C$ must be kept below $\sim 1$.
This can be e.g.~achieved by means of an additional shunt resistor, parallel to the JJ.

\subsection{dc SQUID basics}
\label{subsec:dcSQUID}

\begin{figure}[b]
\includegraphics[width=0.85\columnwidth]{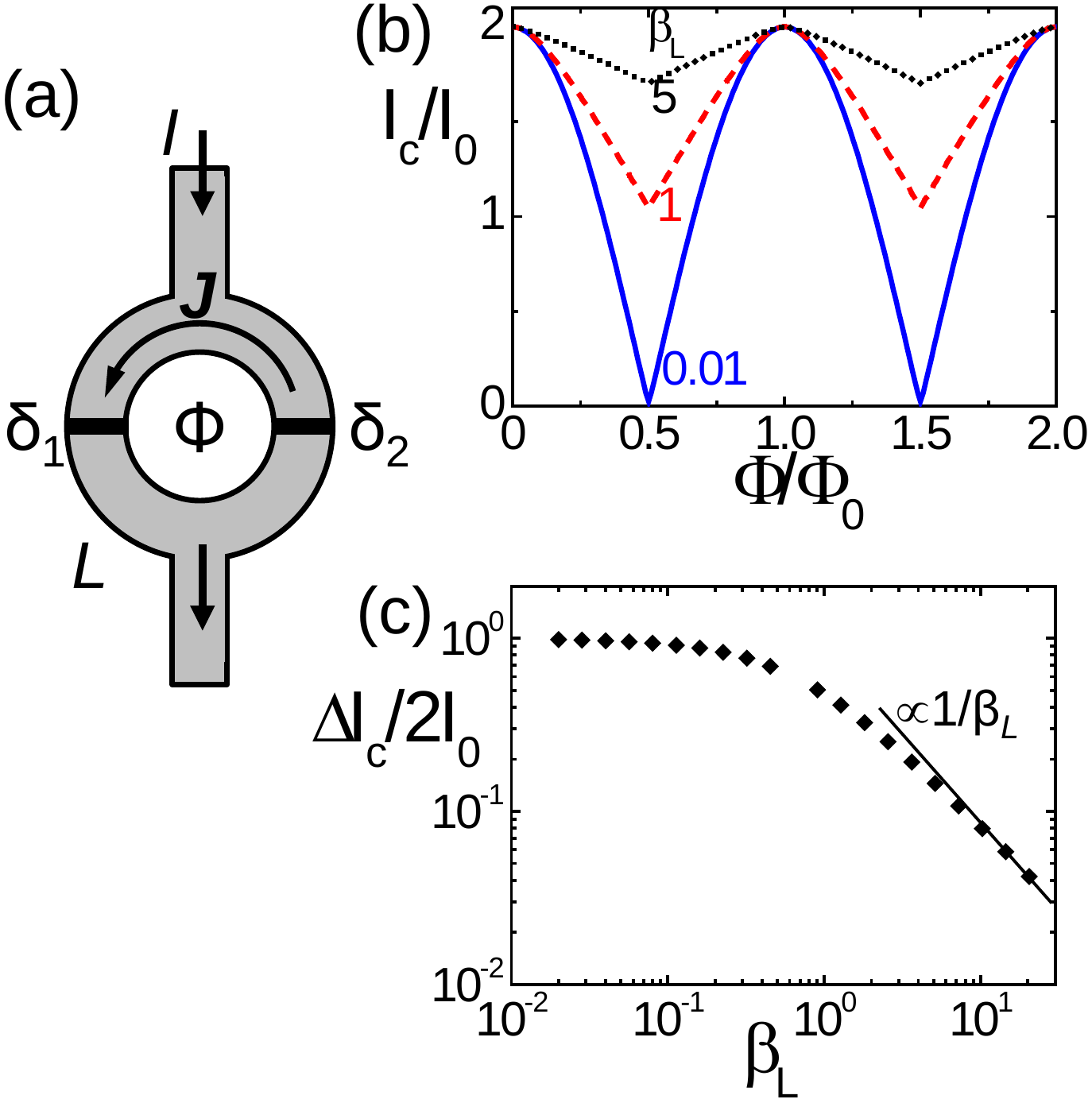}
\caption{The dc SQUID:
(a) Schematic view.
(b) Critical current vs applied magnetic flux for different $\beta_L$ and (c) $I_{\rm c}$ modulation vs $\beta_L$, both calculated for $T=0$ and identical JJs.}
\label{Fig:2}
\end{figure}

The dc SQUID \cite{Jaklevic64} is a superconducting loop (with inductance $L$) intersected by two JJs [Fig.~\ref{Fig:2}(a)].
With an externally applied magnetic flux $\Phi$ through the loop, the fluxoid quantization links the phase differences $\delta_1$ and $\delta_2$ of the two JJs to the total flux in the SQUID $\Phi_{\rm T}=\Phi + LJ$ via
%
\begin{equation}
\delta_1 - \delta_2 + 2\pi n = \frac{2\pi}{\Phi_0} (\Phi + LJ)    \quad  .	
\label{eq:Phi-tot}
\end{equation}
%
Here, $J$ is the current circulating in the SQUID loop and $n$ is an integer \cite{Kleiner-SHB-A1}.
Defining the screening parameter as
%
\begin{equation}
\beta_L \equiv \frac{2 L I_0}{\Phi_0}   \quad  ,	
\label{eq:beta-L}
\end{equation}
%
one finds in the limit $\beta_L\ll 1$ a negligible contribution of $LJ$ to $\Phi_{\rm T}$ in Eq.~(\ref{eq:Phi-tot}), and by assuming for simplicity identical values for $I_0$ in the two JJs, the maximum supercurrent (critical current) $I_{\rm c}$ of the SQUID can be easily obtained as
%
\begin{equation}
I_{\rm c} = 2 I_0 \left|\cos\left(\frac{\pi\Phi}{\Phi_0}\right)\right|   \quad  .
\label{eq:Ic(Phi)}
\end{equation}
%
The pronounced $I_{\rm c}(\Phi)$ dependence  [Fig.~\ref{Fig:2}(b) for $\beta_L\ll 1$] can be used to probe tiny changes in applied magnetic flux.
No analytical expression for $I_{\rm c}(\Phi)$ can be obtained when a finite $\beta_L$ and hence a finite $L$ is included, unless restrictions are imposed to some of the important SQUID parameters \cite{Chesca-SHB-2,Soloviev16}. 
An increasing $\beta_L$ leads to a monotonic decrease of the critical current modulation $\Delta I_{\rm c}/2I_0$ with increasing $\beta_L$  [Fig.~\ref{Fig:2}(b,c)].
This effect allows to estimate $L$ from the measured $I_{\rm c}(\Phi)$.

We note that the inductance $L=L_{\rm g}+L_{\rm k}$ has two contributions \cite{Kleiner-SHB-A1}: The geometric inductance $L_{\rm g}$ relates the induced flux $L_{\rm g} J$ to the current $J$ circulating in the SQUID loop.
The kinetic inductance $L_{\rm k}$ is due to the kinetic energy of $J$ and can often be neglected; however, it becomes significant when the width and/or thickness of the SQUID ring are comparable to or smaller than the London penetration depth $\lambda_{\rm L}$.

For most applications, the dc SQUID is operated in the dissipative state as a flux-to-voltage transducer.
In this case, the SQUID is current biased slightly above $I_{\rm c}$, leading to a $\Phi_0$-periodic modulation of  $V(\Phi)$, which is often sinusoidal.
This mode of operation requires non-hysteretic IVCs, i.e., $\beta_C\lapprox 1$.
An applied flux signal $\delta\Phi$ causes then a change $\delta V$ in SQUID voltage, which for small enough signals is given by $\delta V=(\partial V/\partial\Phi)\,\delta\Phi$.
Usually, the working point (with respect to bias current $I$ and applied bias flux) is chosen such that the slope of the $V(\Phi)$ curve is maximum, which is denoted as the transfer function $V_\Phi =(\partial V/\partial\Phi)_{\rm max}$.

\begin{figure}[b!]
\includegraphics[width=\columnwidth]{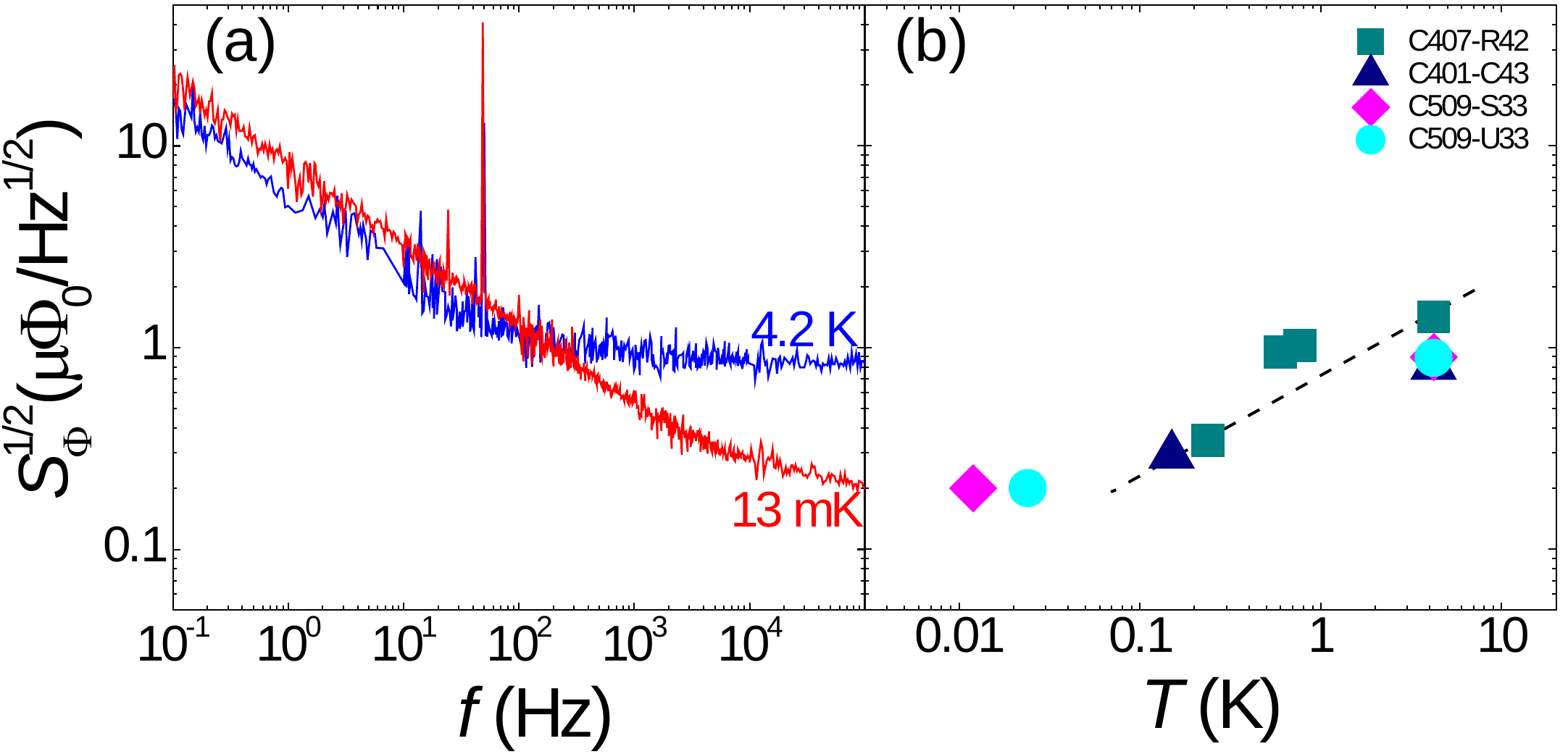}
\caption{Rms flux noise of Nb thin film SQUIDs with Nb/Al-AlO$_x$/Nb JJs
(a) $\sqrt{S_\Phi}(f)$ at 4.2\,K and 13\,mK [after Mart\'{i}nez-P\'{e}rez {\it et al.} \cite{Martinez-Perez10}] 
(b) High-frequency (white) noise, measured at different temperatures on different sensors.
The white noise depends on $T$ as expected from theory ($S_\Phi\propto T$) down to $\sim 100\,$mK when it saturates.}
\label{Fig:3}
\end{figure}

The sensitivity of the SQUID in the voltage state is limited by voltage fluctuations, which are quantified by the spectral density of voltage noise power $S_V$.
This is converted into an equivalent spectral density of flux noise power $S_\Phi=S_V/V_\Phi^2$ or the rms flux noise $\sqrt{S_\Phi}$ with units $\Phi_0/\sqrt{\rm Hz}$ [Fig.~\ref{Fig:3}(a)].

At low frequency $f$, excess noise scaling typically as $S_\Phi\propto 1/f$ (1/$f$ noise) shows up.
Major sources are critical current fluctuations in the JJs and thermally activated hopping of Abrikosov vortices in the superconducting film, which is particularly strong in SQUIDs based on the high-$T_{\rm c}$ cuprate superconductors \cite{Koelle99}.
Moreover, $1/f$ noise has also been ascribed to flux noise arising from fluctuating spins at the interfaces of the devices \cite{Koch07}.
This is supported by the observation of a paramagnetic signal following a Curie-like $T$-dependence \cite{Sendelbach08,Bluhm09,Martinez-Perez11}.
However, a complete description of $1/f$ noise is still missing.

At higher frequencies, $S_\Phi$ becomes indepenent of $f$.
This white noise $S_{\Phi,\rm w}$ is mainly due to Johnson-Nyquist noise associated with dissipative quasiparticle currents in the JJs or shunt resistors.
Within a Langevin approach, the thermal noise is described by two independent fluctuation terms in the coupled equations of motion for the two RCSJ-type JJs.
Numerical simulations yield $S_{\Phi,\rm w}$ vs $\beta_L$, $\beta_C$ and the noise parameter $\Gamma \equiv k_{\rm B} T/E_{\rm J} = 2\pi k_{\rm B} T/(I_0\Phi_0)$ \cite{Koelle99,Chesca-SHB-2}.
For $\beta_C\lapprox 1$, $\beta_L>0.4$ and $\Gamma\beta_L < 0.1$, one finds
%
\begin{equation}
S_\Phi \approx 4(1+\beta_L) \frac{\Phi_0 k_{\rm B} T L}{I_0 R}   
\,.
\label{eq:S-Phi}
\end{equation}
%
For $\beta_L \lapprox 0.4$, $S_\Phi$ increases again with decreasing $\beta_L$.
Typically, SQUIDs are designed to give $\beta_L\approx 1$, for which Eq.~(\ref{eq:S-Phi}) reduces to $S_\Phi\approx 16k_{\rm B}TL^2/R$ \cite{Tesche77}.
This linear scaling $S_\Phi\propto T$, however, saturates in the sub-Kelvin range [Fig.~\ref{Fig:3}(b)] due to the hot-electron effect stemming from limited electron-phonon interaction at low $T$ \cite{Wellstood94}.
We note that $\sqrt{S_\Phi}\propto L$ (for fixed $\beta_L\approx 1$), meaning that small loop inductances yield lower white flux noise levels.
Other sources of white noise are shot and quantum noise, lying usually below the Johnson-Nyquist term.
For the case $\beta_L =1$, the former is given by $S_\Phi\approx hL$ \cite{Tesche77}, whereas the latter arises from zero point quantum fluctuations giving $S_\Phi\approx hL/\pi$ \cite{Koch81}.

\subsection{SQUID readout}
\label{subsec:readout}

\subsubsection{Flux-locked loop}
\label{subsubsec:FLL}

The periodic response of the SQUID to magnetic flux can be linearized to obtain a larger dynamic range.
This can be achieved by operation in the flux locked loop (FLL) mode \cite{Drung-Mueck-SHB-4}.
Here, the SQUID is (typically current) biased at an optimum working point and behaves as a null-detector of magnetic flux.
A small variation $\delta\Phi$ of the external flux changes the SQUID output (typically a voltage change $\delta V$).
This small deviation from the working point is amplified, integrated, and fed back to the SQUID via a current through a feedback resistor $R_{\rm f}$ and  coil, which is inductively coupled to the SQUID.
The output voltage across $R_{\rm f}$ is then proportional to the flux signal $\delta\Phi$.
The dynamic response in FLL mode is limited by the slew rate, i.e.~the speed at which the feedback circuit can compensate for rapid flux changes at the input.
Under optimum conditions, the bandwidth of the FLL is only limited by propagation delays between the room temperature feedback electronics and the SQUID; a typical distance of 1\,m yields $\sim 20\,$MHz.

\subsubsection{Voltage readout}
\label{subsec:V-readout}

The most simple SQUID readout uses current biased operation in the dissipative state; as mentioned above, the IVCs should be non-hysteretic in this case.
%
%
%
%
%
As the transfer function $V_\Phi$ is typically small (several $10-100\,\mu{\rm V}/\Phi_0$), the voltage noise at the output can easily be dominated by room-temperature amplifier noise.
To circumvent this problem, one can use a flux modulation scheme \cite{Drung-Mueck-SHB-4}.
Here, the SQUID is flux-modulated by an ac signal (amplitude $\Phi_0/4$, frequency $f_{\rm m} \sim 100\,$kHz), and the resulting ac voltage across the SQUID is amplified with a (cold) step-up transformer to increase the SQUID signal and noise.
The modulated SQUID response is further amplified at room temperature and lock-in detected.
%
%
%
Suitable electronics achieve a bandwidth of up to 100\,kHz.
%

In a different approach, one can increase $V_\Phi$ by additional positive feedback (APF), which distorts the $V(\Phi)$ characteristics and increases $V_\Phi$ at the positive slope.
This enables simple direct readout of the SQUID signal \cite{Drung-Mueck-SHB-4}. 
Alternatively, a low-noise SQUID or serial SQUID array (SSA) amplifier can be used to amplify the SQUID voltage at low $T$ in a two-stage readout configuration.

\subsubsection{Critical current readout \& threshold detection}
\label{subsubsec:Ic-readout}

For SQUIDs with hysteretic IVCs one can exploit the $I_{\rm c}(\Phi)$ modulation directly.
In this case one ramps the bias current until the SQUID switches to the dissipative state, producing a voltage drop.
At this point the current is switched off, and $I_c$ is calculated from the duration of the ramp \cite{Wernsdorfer09}.
This technique can also be used with a FLL scheme \cite{Wernsdorfer09,Russo12,Granata13a}.
Sensitivity is limited by the accuracy in determining $I_{\rm c}$, 
which is described by the escape of a particle from a potential minimum.
Such a process can be thermally activated or quantum driven and is strongly influenced by electronic noise.
Hence, a large number of switching events is needed to obtain sufficient statistics.


To minimize Joule heating, the SQUID can be operated as a threshold sensor.
Here, the SQUID is current-biased very close to the switching point.
If the magnetic flux threading the loop changes abruptly, 
the SQUID is triggered to the dissipative state and a voltage drop will be measured \cite{Wernsdorfer09}.

Both techniques were applied 
to magnetization reversal measurements on MNPs in sweeping magnetic fields $H$ \cite{Wernsdorfer09}.
For measurements up to large $H$, applied along any direction, the measurement procedure is divided into three steps.
First, $H$ is applied to saturate the particle's magnetization along any direction.
Second, $H$ is swept along the opposite direction to a value $H_{\rm test}$ and back to zero.
To check whether this reversed the particle's magnetization, an in-plane field sweep is done as a third step.
If the particle's magnetization reversal is (not) detected in the third step one can conclude that $H_{\rm test}$ was above (below) the switching field $H_{\rm sw}$.
These steps can be repeated several times to determine $H_{\rm sw}$ precisely.
Note that the second step can be performed above $T_c$ of the SQUID.
Rather than tracing out full $M(H)$ loops, this technique can be used to trace out the dependence of $H_{\rm sw}$ on the field direction and temperature \cite{Jamet01}.

\subsubsection{Dispersive read out}
\label{subsubsec:disp-readout}

So far, we discussed SQUID operation in the voltage state or close to it.
Such schemes entail dissipation of Joule power that might affect the state of the magnetic system under study.
An elegant way to circumvent this problem is the operation of the SQUID as a flux-dependent resonator; this has also the advantage of increasing enormously the bandwidth up to $\sim 100\,$MHz \cite{Hatridge11,Levenson-Falk13}.
The SQUID is always in the superconducting state and acts as a flux-dependent inductance connected in parallel to a capacitor.
The resonance frequency of the circuit depends on the total flux threading the SQUID loop.
This can be read out by conventional microwave reflectometry giving a direct flux-to-reflected phase conversion.
The devices are operated in the linear regime, i.e., using low-power driving signals.
To determine the spectral density of flux noise, the overall voltage noise of the circuit is estimated and scaled with the transduction factor ${\rm d}V/{\rm d}\Phi$.
%
The noise performance can be boosted considerably by taking advantage of the CPR non-linearity, i.e., operating the nanoSQUID as a parametric amplifier.
For this purpose, the driving power is increased so that the resonance peak is distorted, giving a much sharper dependence of the reflected phase on $\Phi$.
%


\section{nanoSQUIDs: design, fabrication \& performance}
\label{sec:nanoSQUIDs}

\begin{figure}[t]
\includegraphics[width=\columnwidth]{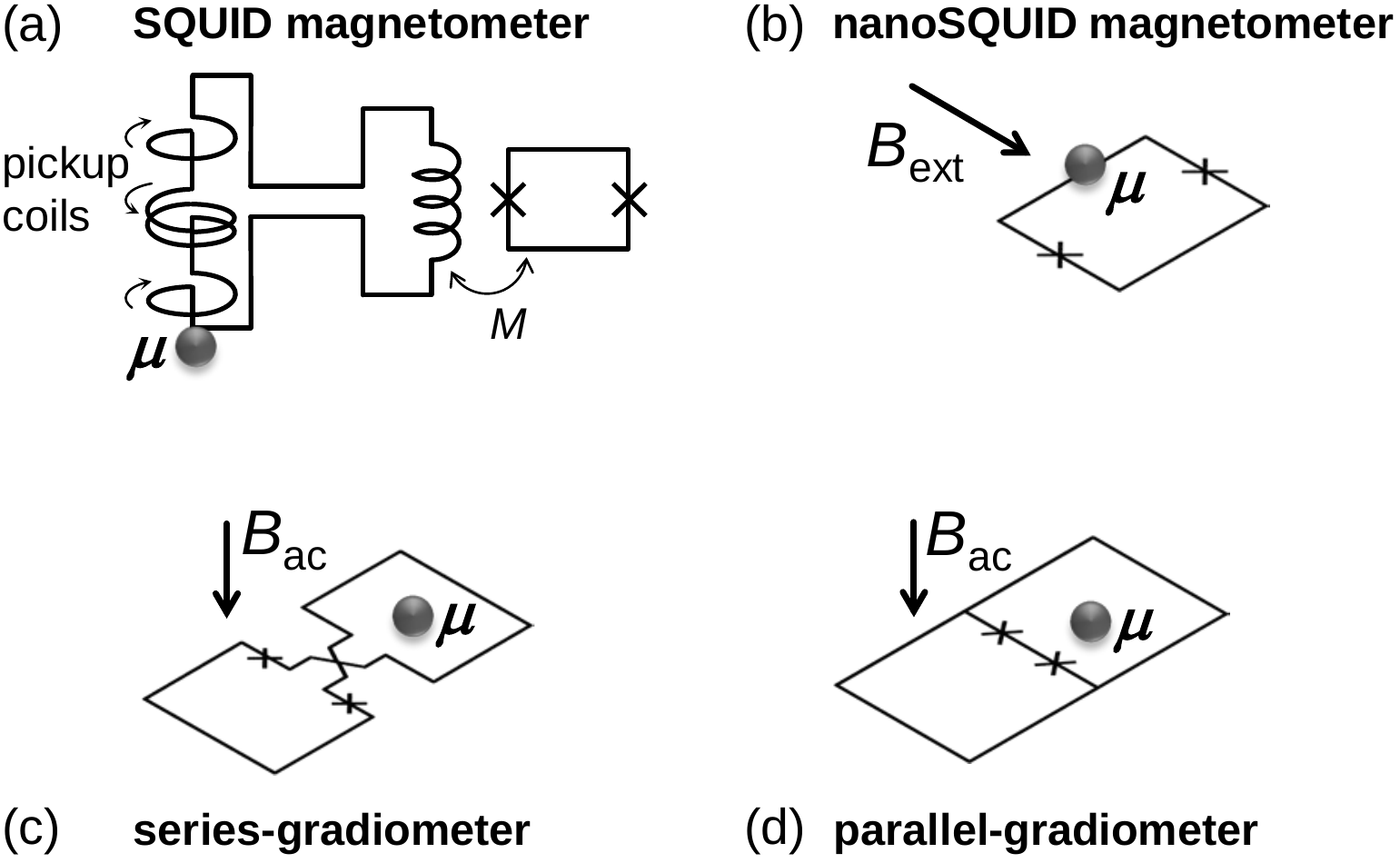}
\caption{Layouts of various SQUID sensors.
(a) SQUID magnetometer based on gradiometric pickup coils coupled inductively (via mutual inductance $M$) to a SQUID.
(b)-(d) NanoSQUIDs without intermediate pick-up coils; the stray field created by a MNP with magnetic moment $\mu$ is directly sensed by the SQUID loop.
Magnetization measurements can be performed by applying an external magnetic field $B_{\rm ext}$ in the nanoloop plane (b).
The frequency-dependent magnetic ac susceptibility $\chi_{\rm ac}$ can be sensed by using series (c) or parallel (d) planar gradiometers; a homogeneous ac excitation magnetic field $B_{\rm ac}$ is applied perpendicular to the gradiometer's plane through on-chip excitation coils. }
\label{Fig:4}
\end{figure}

NanoSQUIDs are developed for detecting small spin systems, such as MNPs or SMMs, or for high-resolution imaging of magnetic field structures by SQUID microscopy.
For such applications, the figure of merit is the spin sensitivity, which can be boosted down to the level of a single electron spin.
%
%
The use of strongly miniaturized SQUID loops and JJs is based on the following ideas:
\begin{itemize}
\item[$\bullet$]
Stronlgy localized magnetic field sources (e.g.~MNPs) are placed in close vicinity to the SQUID, instead of using pickup coils [Fig.~\ref{Fig:4}(a)] which degrade the overall coupling.
A single SQUID loop [Fig.~\ref{Fig:4}(b)] can be used to detect the magnetic moment $\mu$ of a MNP, or gradiometric configurations  [Fig.~\ref{Fig:4}(c,d)] enable measurements of the magnetic ac susceptibility $\chi_{\rm ac}$.
\item[$\bullet$]
The coupling of the stray field from local field sources to the nearby SQUID can be improved by reducing the cross section (width and thickness) of the superconducting thin film forming the SQUID loop  (see section \ref{subsec:design}). 
\item[$\bullet$]
The sensitivity of the SQUID to magnetic flux (magnetic flux noise in the thermal white noise limit) can be improved by reducing the loop inductance, i.e.~by shrinking the lateral size of the SQUID loop (see section \ref{subsec:design}). 
\item[$\bullet$]
For magnetization reversal measurements on MNPs, an external field $B_{\rm ext}$ is applied ideally exactly in the plane  of the SQUID loop to switch the MNP's magnetization (see section \ref{subsec:Magnetization}), albeit without coupling flux directly to the loop.
By reducing the dimensions of the JJs and the loop, the nanoSQUID can be made less sensitive to $B_{\rm ext}$ for small misalignment of $B_{\rm ext}$. 
\item[$\bullet$]
Reducing the loop size together with the SQUID-to-sample distance can significantly boost the spatial resolution for scanning SQUID microscopy applications (see section \ref{sec:SSM}).
\end{itemize}

\subsection{nanoSQUIDs: design considerations}
\label{subsec:design}

The ability of a nanoSQUID to resolve tiny signals from the magnetic moments of small spin systems depends (i) on the intrinsic flux noise $S_\Phi$ of the SQUID and (ii) on the amount of flux $\Phi$ which a particle with magnetic moment $\bm\mu$ couples to the SQUID loop.
The latter can be quantified by the coupling factor $\phi_\mu \equiv \Phi/\mu$, with $\mu\equiv |\bm\mu|$.
As a result, one can define the spin sensitivity $\sqrt{S_\mu} =\sqrt{S_\Phi}/\phi_\mu$, with units $\mu_{\rm B}/\sqrt{\rm Hz}$; $\mu_{\rm B}$ is the Bohr magneton.
$\sqrt{S_\mu}$ expresses the minimum magnetic moment that can be resolved per unit bandwidth.
Hence, optimizing nanoSQUID performance requires to minimize $S_\Phi$ while maximizing $\phi_\mu$.

As mentioned in section \ref{subsec:dcSQUID}, $S_\Phi$ has typically a low-frequency 1/$f$-like contribution and a thermal white noise part $S_{\Phi, \rm w}$.
The 1/$f$ contribution is hard to optimize by design; however, $S_{\Phi, \rm w}$ depends on geometrical parameters through the loop inductance $L$, but also on junction parameters such as $I_0$, $R$ and $C$.
The $S_\Phi(L)$ dependence (Eq.(\ref{eq:S-Phi})) implies that $S_\Phi$ can be improved by decreasing $L$ via the loop dimensions, while considering the constraints on $\beta_C$ and $\beta_L$, which will affect the choice of junction parameters.
Such an optimization procedure can be tested experimentally by performing flux noise measurements of the SQUIDs.

The optimization of the coupling factor $\phi_\mu = \Phi/\mu$ is more difficult.
It is defined as the magnetic flux $\Phi$ coupled to the SQUID loop by the magnetic dipole field of a point-like particle, divided by its magnetic moment $\mu$.
The magnitude of $\phi_\mu$ depends on SQUID geometry, particle position $\bm r_\mu$ (relative to the SQUID) and orientation $\hat{\bm e}_\mu=\bm\mu/\mu$ of its magnetic moment.
This quantity is not directly accessible by experiments, and one has to rely on estimates, analytic approximations or numerical calculations for determining $\phi_\mu$ and  optimizing it.

To the best of our knowledge, Ketchen et al. \cite{Ketchen89} were the first to give an estimate of $\phi_\mu$.
For a magnetic dipole at the center of an infinitely thin loop with radius $a$, with $\hat{\bm e}_\mu$ along the loop normal
\begin{equation}
\phi_\mu=\frac{\mu_0}{2a} 
=
(r_e/a)\cdot(\Phi_0/\mu_{\rm B})
\approx
(2.8\,\mu{\rm m}/a)\cdot({\rm n}\Phi_0/\mu_{\rm B})
\label{eq:phi-mu-Ketchen}
\end{equation}
was found.
%
\footnote{$\phi_\mu=2\pi/a$ in cgs units, as derived by Ketchen et al.~\cite{Ketchen89}.
The spin sensitivity $S_n$ in \cite{Ketchen89} relates to our definition as $S_n = \sqrt{S_\mu}/\mu_{\rm B}$, i.e.~$S_n$ has the units of number of spins (of moment $\mu_{\rm B}$) per $\sqrt{\rm Hz}$.}
%
%
%
The r.h.s.~of Eq.~(\ref{eq:phi-mu-Ketchen}) is obtained  with the definition of the classical electron radius $r_e=\frac{\mu_0e^2}{4\pi m_e}$, $\Phi_0=\frac{h}{2e}$ and $\mu_{\rm B}=\frac{eh}{4\pi m_e}$, which yields $\frac{\mu_{\rm B}}{\Phi_0}=\frac{2r_e}{\mu_0}$.
%

The coupling improves if the particle is moved close to the loop's banks \cite{Bouchiat09}.
However, a quantitative estimate of $\phi_\mu$ is more difficult in this near-field regime \cite{Tilbrook09}, as the cross-section of the SQUID banks and the flux focusing effect caused by the superconductor must be taken into account.
The calculation of $\phi_\mu$ requires calculating the magnetic field distribution at the position of the SQUID, originating from a magnetic moment $\bm\mu$ at position $\bm r_\mu$, and from this the magnetic flux coupled to the SQUID.
This problem can be simplified by exploiting the fact that sources and fields can be interchanged, i.e., one evaluates the magnetic field $\bm B_J (\bm r_\mu)$, created by a circulating supercurrent $J$ through the SQUID loop, at the position $\bm r_\mu$ of the magnetic dipole.
With the normalized quantity $\bm b_J = \bm B_J/J$, which does not depend on $J$, one finds \cite{Bouchiat09,Nagel11}
%
\begin{equation}
\phi_\mu (\bm r_\mu,\hat{\bm e}_\mu) = \hat{\bm e}_\mu \cdot \bm b_J(\bm r_\mu) \quad .
\label{eq:phi-mu}
\end{equation}
%
This allows to calculate $\phi_\mu$ for any position and orientation of the magnetic dipole in 3D space once $\bm b_J$ is known.
%
\footnote{The current $J$ through an infinitely thin wire, forming a loop with radius $a$ in the $x-y$ plane and centered at the origin, induces a field $\bm B_J=\mu_0 J/(2a) \cdot \hat{\bm e}_z$, at the center of the loop.
Hence, for a magnetic dipole placed at the origin $\bm r_\mu=0$ and pointing in $z$ direction, $\hat{\bm e}_\mu = \hat{\bm e}_z$, Eq.~(\ref{eq:phi-mu}) yields $\phi_\mu = \hat{\bm e}_z \cdot \bm B_J(\bm r_\mu)/J=\mu_0/(2a)$, i.e. the same result as derived by Ketchen et al. \cite{Ketchen89}.}

\begin{figure}[t]
\includegraphics[width=\columnwidth]{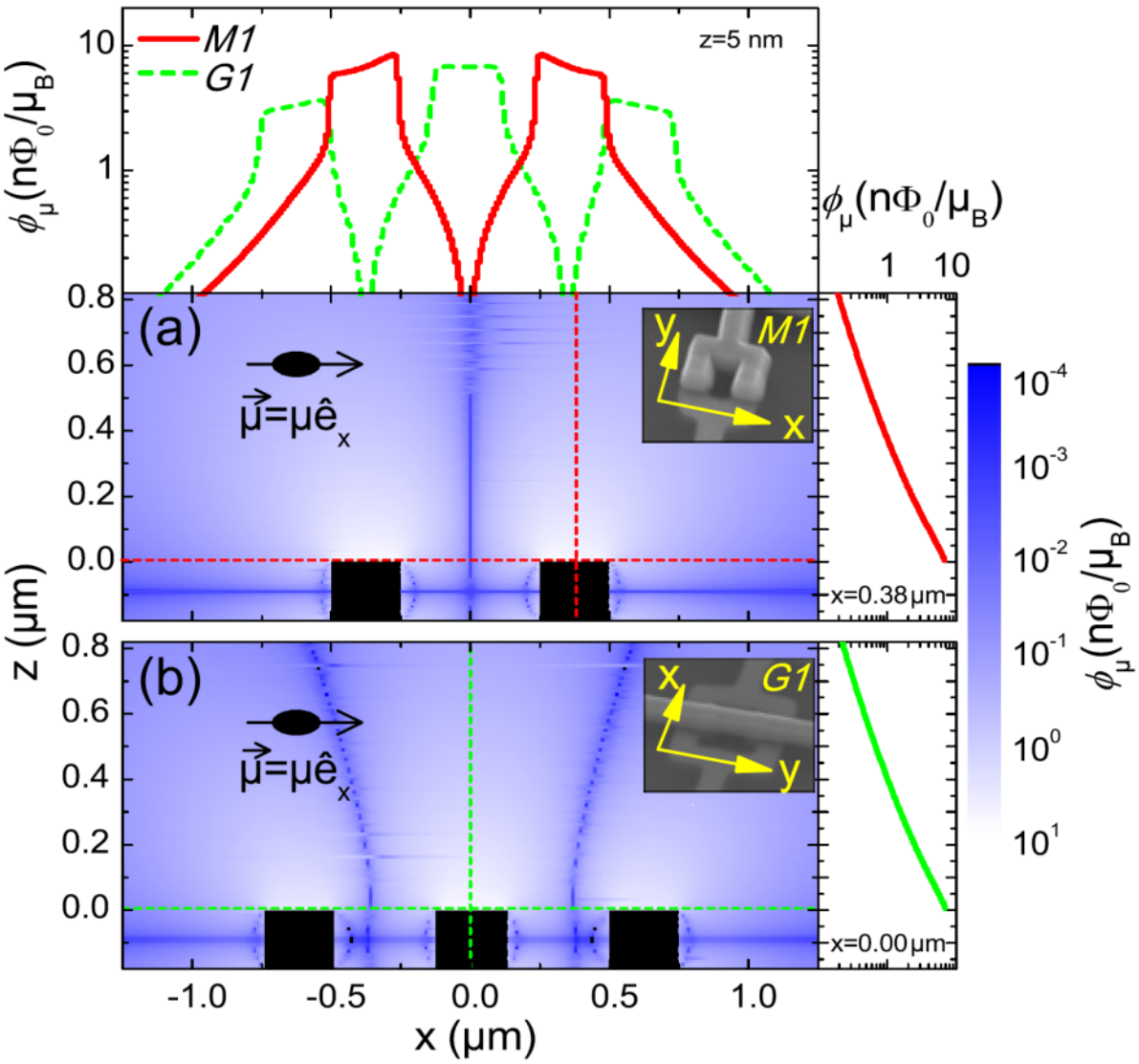}
%
\caption{Calculated coupling factor $\phi_\mu$ vs position of a magnetic dipole pointing in $x$-direction on top of Nb nanoSQUIDs.
Main graphs show contour plots $\phi_\mu(x,z)$ for (a) a magnetometer and (b) a gradiometer.
Nb structures are indicated by black rectangles; dashed lines indicate position of linescans $\phi_\mu(x)$ [above (a)] and $\phi_\mu(z)$ [right graphs].
Insets show scanning electron microscopy (SEM) images.
Reprinted with permission from Nagel {\it et al.}\cite{Nagel11a}.
Copyright [2011], AIP Publishing LLC.}
\label{Fig:5}
%
\end{figure}

The normalized field $\bm b_J$ has to be determined from the spatial distribution of the supercurrent density $j_{\rm s}$ circulating in the SQUID loop, which depends only on the SQUID geometry and on $\lambda_{\rm L}$. 
This has been done for various types of nanoSQUIDs by numerically solving the London equations
\cite{Nagel11,Nagel11a,Nagel13,Schwarz13,Schwarz15,Woelbing13,Woelbing14}.
Numerical simulations of $\phi_\mu$ reveal that the coupling can be increased in the near-field regime if the magnetic dipole is placed as close as possible on top of a constriction in the SQUID loop, which is as thin and narrow as possible \cite{Woelbing14}.
Typical $\phi_\mu=10-20\,{\rm n}\Phi_0/\mu_{\rm B}$ have been obtained for magnetic dipoles at 10\,nm distance from a constriction ($\sim 100-200\,$nm wide and thick) in YBa$_2$Cu$_3$O$_7$ (YBCO) nanoSQUIDs.
%
\footnote{$\phi_\mu$ depends significantly on the loop width, thickness $d$ and $\lambda_{\rm L}$.
For example for a dipole centered at a circular SQUID loop with inner radius $a=500\,$nm, outer radius $R=2\,\mu$m, and $d=\lambda_{\rm L}=100\,$nm one finds $\phi_\mu=3.5\,{\rm n}\Phi_0/\mu_{\rm B}$, i.e. a factor 1.6 smaller $\phi_\mu$ as obtained from Ref.~\cite{Ketchen89} (with $R=a=500\,$nm); $\phi_\mu$ decreases further with decreasing ratio $d/\lambda_{\rm L}$.} 
%
Simulation results for two types of Nb nanoSQUIDs [Fig.~\ref{Fig:5}] show that the dipole has to approach the SQUID surface closely to reach values above a few n$\Phi_0/\mu_{\rm B}$ (see $\phi_\mu(z)$ linescans in the right graphs in Fig.~\ref{Fig:5}).
The $\phi_\mu(x)$ linescans (top graph in Fig.~\ref{Fig:5}) show that the coupling is maximum right above the loop structures \cite{Nagel11a}.

Measurements on spatially extended magnetic systems, such as a Ni nanotube \cite{Nagel13} or a Fe nanowire \cite{Schwarz15}, were found  to be consistent with the numerical approach described above.
This was done by comparing the measured flux coupled to nanoSQUIDs from fully saturated tubes or wires with the calculated flux signals, obtained by integrating $\phi_\mu$ over the finite volume of the sample.
First measurements on the SQUID response as a function of the position of a magnetic sample have been reported earlier.
In those experiments, small SQUID sensors were coupled to a ferromagnetic Fe tip, which was scanned over the sensor's surface while recording the SQUID output in open-loop configuration \cite{Josephs-Franks03}.

The optimization of the spin sensitivity in the thermal white noise limit requires the knowledge of the dependence of $\phi_\mu$ and $S_{\Phi, {\rm w}}$ on SQUID geometry, as this affects both the SQUID inductance and the coupling.
A detailed investigation of this problem was done for YBCO nanoSQUIDs \cite{Woelbing14} (see section \ref{subsec:high-Tc}).
This study shows that it is essential to consider the increase in kinetic inductance $L_{\rm k}$ when the thickness and width of the loop is reduced to a length scale comparable to or even smaller than $\lambda_{\rm L}$.
Hence, to improve the $S_\mu$ one has to find a compromise between improved coupling and deterioration of flux noise (via an increased $L_{\rm k}$) upon shrinking the cross section of the SQUID loop.

\subsection{nanoSQUIDs based on metallic superconductors}
\label{subsec:low-Tc}


\subsubsection{Sandwich-type SIS junctions}
\label{subsubsec:SIS-JJs}

The SIS junction technology (S: superconductor, I: thin insulating barrier), typically producing JJs in a Nb/Al-AlO$_x$/Nb trilayer geometry, is the most commonly used approach to fabricate conventional SQUID-based devices.
This technology is highly developed and reproducible, yielding high-quality JJs with controllable critical current densities $j_{\rm c}$ from $\sim 0.1$ up to a few kA/cm$^2$ at 4.2\,K.
However, a major disadvantage is the low $j_{\rm c}$, which results in too small values for the critical current if submicron JJs are used.
As a consequence, even if the SQUID loops are miniaturized, the operation of micron-sized JJs in large magnetic fields is only possible with careful alignment of the field perpendicular to the junction plane, as an in-plane field in the 1-10\,mT range can easily suppress the critical current due to the Fraunhofer-like modulation of $I_{\rm c}(B)$.
Frequently used window-type JJs come with a large parasitic capacitance due to the large area of surrounding superconducting layers.
A commonly used approach is therefore to use normal metal layers to shunt these junctions, for lowering $\beta_C$ to yield non-hysteretic IVCs, albeit at the cost of also lowering the characteristic voltage $V_{\rm c} = I_0R$.
The absence of hysteresis offers the advantage to operate the SQUID as a flux-to-voltage converter, using conventional readout techniques.

As a key advantage, the SIS technology offers a well developed multilayer process, allowing for the realization of more complex designs, as compared to a single layer technology.
This allows for the fabrication of superconducting on-chip input circuits such as coupling transformers, susceptometers or advanced gradiometers.
This approach has been taken very successfully to realize miniaturized structures for applications in magnetic particle measurements and scanning SQUID microscopy, although those did not really involve SQUIDs  with (lateral outer) dimensions in the submicrometer range.

\begin{figure*}[t!]
\includegraphics[width=0.8\textwidth]{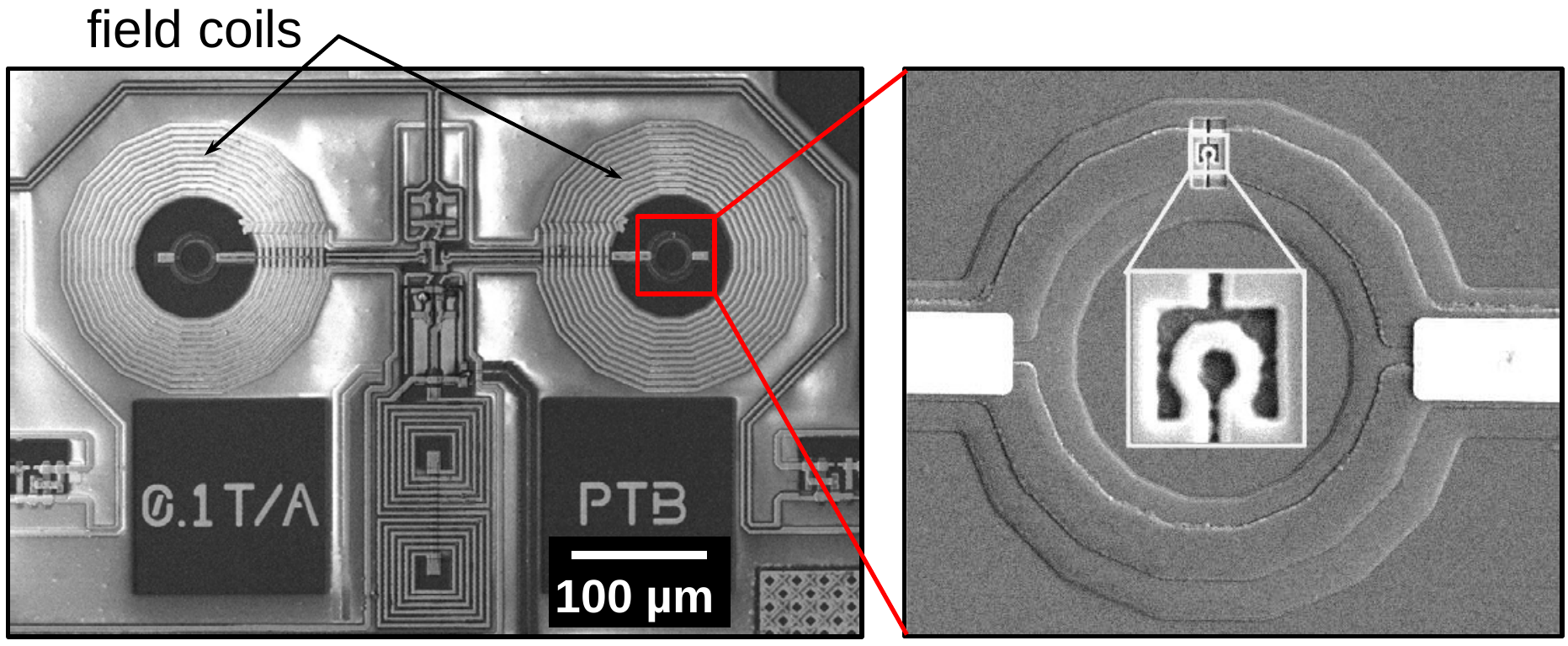}
\hastfill
\begin{minipage}[b]{0.18\textwidth}
\caption{SEM image of a SQUID microsusceptometer which a nanoloop patterned in the pickup coil (inset).
Images courtesy of J. Ses\'{e}.}
\label{Fig:6}
\end{minipage}
\end{figure*}

The first SQUID device designed to measure magnetic signals from MNPs was based on micrometric Nb/NbO$_x$/Pb edge junctions,
which were connected in parallel to two oppositely wound loops to form a micro-susceptometer \cite{Ketchen84}.
The white flux noise 
at 4.2\,K was $0.84\,\mu\Phi_0/\sqrt{\rm Hz}$.
This susceptometer was operated in a dilution refrigerator, and the output signal was measured in open-loop configuration and amplified by an rf SQUID preamplifier.
Magnetic susceptibility measurements performed with this system will be reviewed in section \ref{subsec:Susceptibility}.
Very similar devices based on Nb/Al-AlO$_x$/Nb JJs with $\sqrt{S_\Phi}=0.8\,\mu\Phi_0/\sqrt{\rm Hz}$ at 4\,K and $0.25\,\mu\Phi_0/\sqrt{\rm Hz}$ below 0.5\,K were adapted to the use in scanning SQUID microscopes \cite{Gardner01,Huber08}; see section \ref{sec:SSM}.

Broad-band SQUID microsusceptometers have been realized by locally modifying SQUID current sensors 
based on Nb/Al-AlO$_x$/Nb JJ technology.
Those sensors \cite{Drung07}
%
%
%
%
come in two types: (i) high-input inductance ($\sim 1\,\mu$H) sensors incorporate an intermediate transformer loop with gradiometric design; (ii) low-input inductance (2\,nH) devices without intermediate loop; here the input signal is directly coupled to the SQUID via four single-turn gradiometric coils connected in parallel.
These SQUIDs are non-hysteretic down to sub-K temperatures 
with $\sqrt{S_{\Phi, \rm w}}=800\,{\rm n}\Phi_0/\sqrt{\rm Hz}$ at $T = 4.2\,$K.
Modification of these sensors was done by FIB milling and FIB-induced deposition (FIBID) of superconducting material with W(CO)$_6$ as precursor gas \cite{Martinez-Perez09,Martinez-Perez10}.
This allowed converting the intermediate transformer loop into a susceptometer inductively coupled to the SQUID [Fig.~\ref{Fig:4}(a)].
By modifying the gradiometric microSQUID itself it is possible to directly couple an MNP to the SQUID loop \cite{Martinez-Perez11} [Fig.~\ref{Fig:4}(d)].
Later, SQUID-based microsusceptometers with improved reflection symmetry were produced \cite{Drung14,Schurig14}.
The sensitivity was boosted by defining a nanoloop (450\,nm inner diameter, 250\,nm linewidth) by FIB milling in one of the pickup coils [Fig.~\ref{Fig:6}].
These sensors offer an extremely wide bandwidth (1\,mHz - 1\,MHz) and can be operated at $T=0.013 - 5\,$K for the investigation of microscopic crystals of SMMs and magnetic proteins; such measurements will be reviewed in section \ref{subsec:Susceptibility}.

Submicrometric Nb/AlO$_x$/Nb JJs in a cross-type design were recently used for fabricating miniaturized SQUIDs \cite{Schmelz12}.
The key advantage of cross-type JJs over conventional window-type JJs is the elimination of the parasitic capacitance surrounding the JJ, which becomes increasingly important upon reducing the JJ size.
At $T=4.2\,$K, $0.8\times 0.8\,\mu{\rm m}^2$ JJs 
show non-hysteretic IVCs, if they are shunted with a AuPd layer.
%
Sensors are also produced with an integrated Nb modulation coil.
Square-shaped washer SQUIDs with minimum inner size of $0.5\,\mu$m have an inductance of a few pH.
%
SQUIDs operated in liquid He and read out with a low-noise SQUID preamplifier yield $\sqrt{S_{\Phi,\rm w}}=66\,{\rm n}\Phi_0/\sqrt{\rm Hz}$ \cite{Schmelz15}.

\subsubsection{Sandwich-type SNS junctions}
\label{subsubsec:SNS-JJs}

SNS junctions (N: normal conductor) offer the advantage of large critical current densities $\gapprox 10^5\,{\rm A/cm}^2$ at 4.2\,K and non-hysteretic IVCs, 
albeit at the cost of somewhat reduced $I_0R$ values.
Hence, this type of JJs is very well suited for fabricating nanoSQUIDs with junction size in the deep submicron range.

\begin{figure}[t]
\includegraphics[width=0.9\columnwidth]{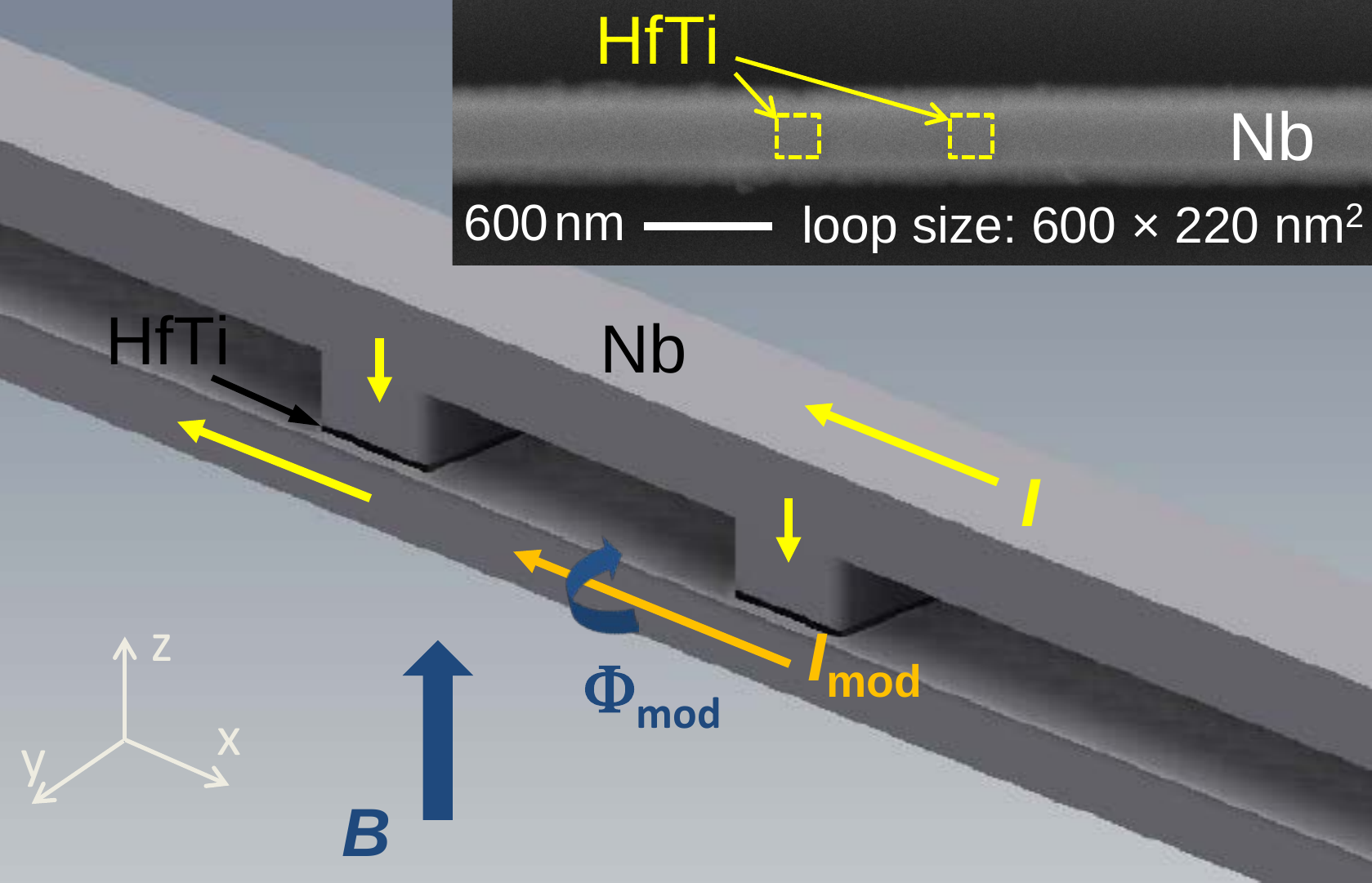}
%
\caption{Layout of Nb/HfTi/Nb nanoSQUID in microstrip geometry.
Arrows indicate flow of bias current $I$, modulation current $I_{\rm mod}$ and direction of external field $B$.
Inset shows SEM image with JJs ($200 \times 200\,{\rm nm}^2$) indicated by dashed squares.
SEM image courtesy of B.~M\"{u}ller.}
\label{Fig:7}
%
\end{figure}

In a Nb/HfTi/Nb trilayer process, originally developed for Josephson arbitrary waveform synthesizers \cite{Hagedorn06}, JJs with $200 \times 200\,{\rm nm}^2$ area or even below are obtained by e-beam lithography and chemical-mechanical polishing, producing nanoSQUIDs \cite{Nagel11a,Woelbing13} 
with 24\,nm thick HfTi barriers; the latter can be varied to modify $j_c$.
%
%
As for the SIS JJ technology, the fabrication process offers much flexibility for realizing complex designs.
Both series- and parallel-gradiometers and single SQUID loops were realized \cite{Nagel11a,Woelbing13,Bechstein15}.
Devices were patterned in a washer- 
or microstrip-type geometry, 
with the loop plane parallel or perpendicular to the junction's (substrate) plane, respectively.
A key advantage of the microstrip-type geometry [Fig.~\ref{Fig:7}] is the possibility to realize very small loop areas, defined by the thickness of the insulating interlayer between the top an bottom Nb lines times the lateral separation of the two JJs.
This results in very small SQUID inductances, typically a few pH. 
Moreover, a magnetic field applied in the plane of the SQUID loop can be perpendicular to the JJ (and substrate) plane; in this way the field-induced suppression of $I_{\rm c}$ can be avoided.
It has been shown that magnetic fields up to 0.5\,T can be applied while degrading only marginally the performance \cite{Woelbing13}.
On-chip flux biasing is easily possible for operation in FLL.
%
%
%
White flux noise $\sim 110\,{\rm n}\Phi_0/\sqrt{\rm Hz}$ has been obtained.
Based on numerical solutions of the London equations for $\phi_\mu$, this yields a spin sensitivity of just $\sim 10\,\mu_{\rm B}/\sqrt{\rm Hz}$ for a magnetic dipole 10\,nm away from the SQUID loop.
Magnetization measurements on magnetic nanotubes have been performed successfully and will be summarized in section \ref{subsec:Magnetization}.

\begin{figure}[b!]
\includegraphics[width=0.9\columnwidth]{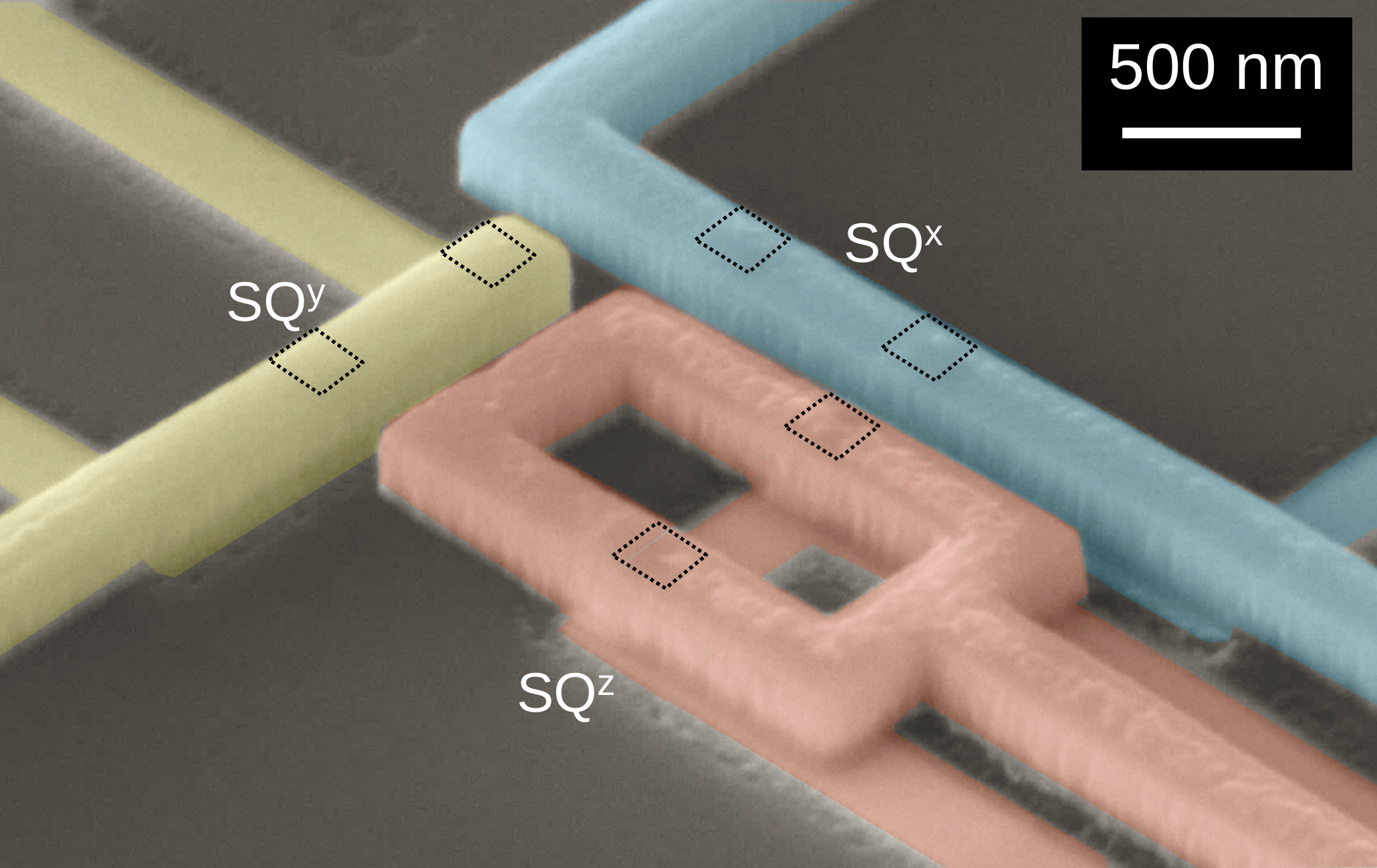}
%
\caption{SEM image of a 3-axis vector magnetometer, consisting of two orthogonal nanoSQUIDs (SQ$^x$, SQ$^y$) and an orthogonal gradiometric nanoSQUID (SQ$^z$).
Black dotted squares indicate  positions of Josephson junctions.}
\label{Fig:8}
%
\end{figure}

By combining three mutually orthogonal nanoSQUID loops, a 3-axis vector magnetometer has been realized very recently \cite{Martinez-Perez16}.
Here, the idea is to distinguish the three components of the vector magnetic moment $\bm \mu$ of a MNP placed at a specific position, and subjected to an applied magnetic field along $z$-direction for magnetization reversal measurements.
The layout of the device is shown in Fig.~\ref{Fig:8}.
Two microstrip-type Nb nanoSQUIDs SQ$^x$ and SQ$^y$, as described above, with perpendicular loops are sensitive to fields in $x$- and $y$-direction, respectively.
A third SQUID, SQ$^z$ has a gradiometric layout, in order to strongly reduce its sensitivity to the applied homogeneous magnetic field.
Simultaneous operation of all three nanoSQUIDs in such devices in FLL has been demonstrated at 4.2\,K in fields up to 50\,mT, with a flux noise $S_{\Phi, \rm w}^{1/2}\lapprox 250\,{\rm n}\Phi_0/\sqrt{\rm Hz}$.
By numerical simulations of the coupling factor, it has been demonstrated that for a MNP placed in the center of the left loop of the gradiometer (cf.~Fig.~\ref{Fig:8}), the three orthogonal components of the magnetic moment of the MNP can be detected with a relative error flux below 10\,\%.
Such a device can provide important information on the magnetic anisotropy of a single MNP.

\begin{figure}[t]
\includegraphics[width=0.9\columnwidth]{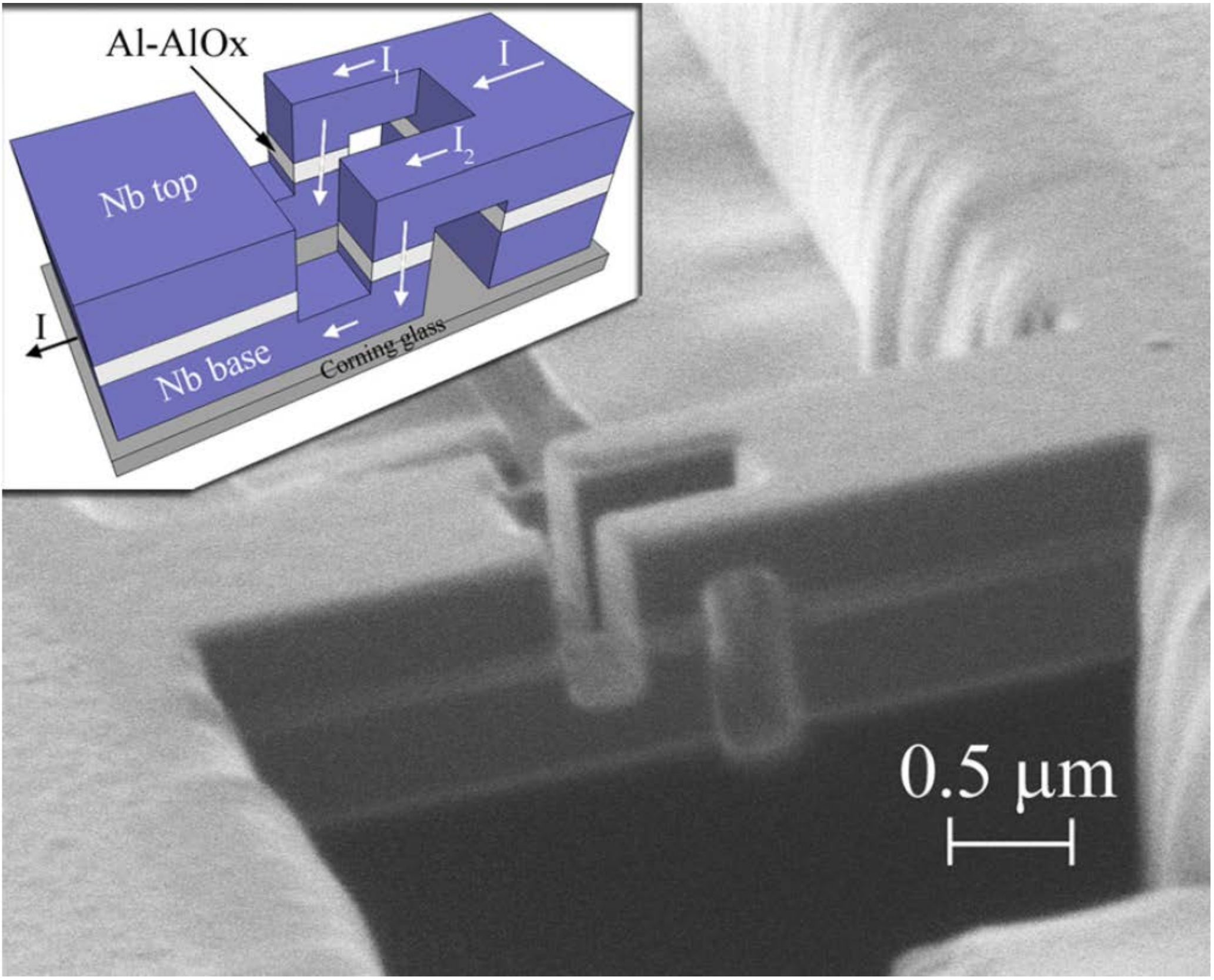}
%
\caption{SEM image of a 3-dimensional nanoSQUID fabricated using FIB sculpting and all Nb technology.
The flux capture area of the nanosensor is $1\times 0.2\,\mu{\rm m}^2$, and the two Josephson tunnel junctions have an area of about $0.3\times 0.3\,\mu{\rm m}^2$.
The inset is a sketch of the device, showing the current paths through the device.
Reprinted with permission from Granata {\it et al.} \cite{Granata13}.
Copyright [2013], AIP Publishing  LLC.}
\label{Fig:9}
%
\end{figure}

Submicrometer nanoSQUIDs have recently also been fabricated based on SNIS JJs \cite{Granata13}.
Starting from a Nb/Al-AlO$_x$/Nb trilayer, a three dimensional SQUID loop ($0.2\,\mu{\rm m}^2$) was nanopatterned by FIB milling and anodization [Fig.~\ref{Fig:9}].
The resulting JJs have an area of approximately $0.3\times 0.3\,\mu{\rm m}^2$ and are intrinsically shunted by the relatively thick (80\,nm) Al layer, yielding non-hysteretic IVCs.
%
The smallness of the SQUID loop leads to $L=7\,$pH.
%
Measurements at 4.2\,K yield 
$\sqrt{S_{\Phi, \rm w}}\sim 0.68\,\mu\Phi_0/\sqrt{\rm Hz}$.

\subsubsection{Constriction junctions}
\label{subsubsec:cJJs}

\begin{figure*}[t]
\includegraphics[width=0.73\textwidth]{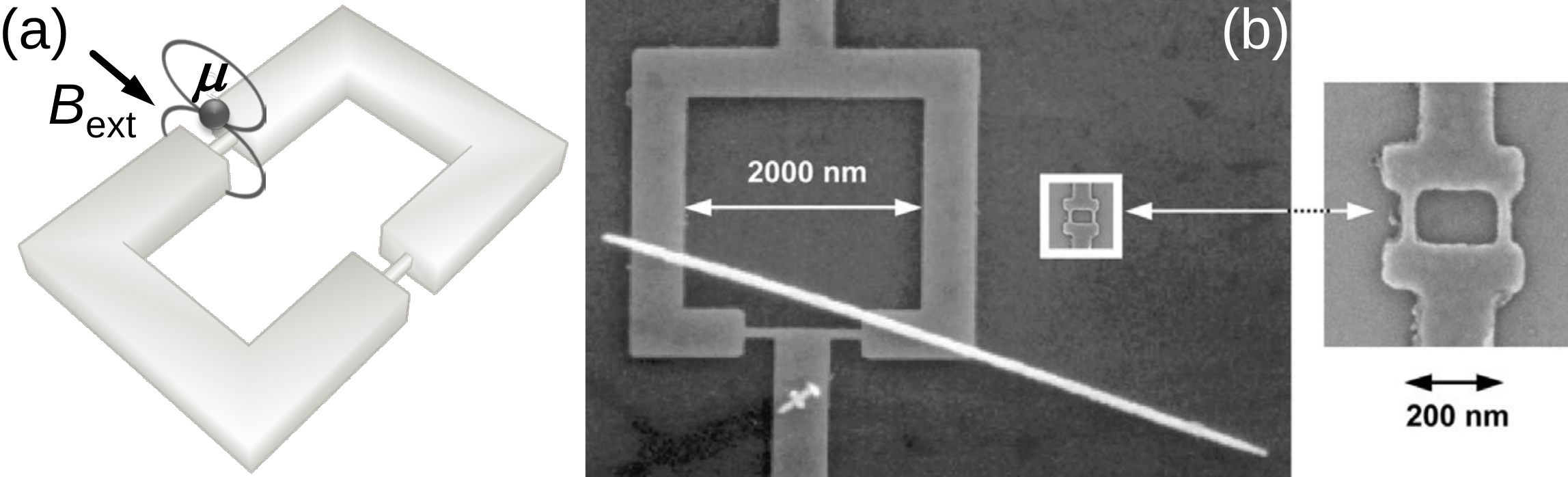}
\hastfill
\begin{minipage}[b]{0.25\textwidth}
\caption{cJJ-based nanoSQUIDs:
(a) schematic view with a MNP (magnetic moment $\mu$) close to one constriction where coupling is maximum.
(b) SEM images of Nb microSQUID with Ni wire on top (left) and Nb nanoSQUID (right), drawn to scale in left graph.
Graph (b) Reproduced with permission from \cite{Wernsdorfer09}.
All rights reserved \copyright\,IOP Publishing [2009].}
\label{Fig:10}
\end{minipage}
\end{figure*}

Josephson coupling can also occur in superconducting constrictions (Dayem bridges \cite{Anderson64}) with size similar to or smaller than the coherence length $\xi(T)$ \cite{Likharev79}.
The IVCs of such constriction-type Josephson junctions (cJJs) are often hysteretic, due to the heat dissipated above $I_{\rm c}$.
Short enough cJJs show a sinusoidal CPR; however, a significant deviation occurs if the constriction length is larger than $\xi$, which can even lead to multivalued CPRs.
Hence, optimization of SQUID performance based on an RCSJ analysis is difficult, and hysteretic IVCs prevents conventional SQUID operation with current bias.
Still, non-hysteretic IVCs can be achieved by operation close enough to $T_{\rm c}$, were $I_{\rm c}$ is reduced, or by adding a metallic overlayer as a resistive shunt.
%
Another drawback is the large kinetic inductance $L_{\rm kin}$ of the constriction, that can dominate the total SQUID inductance $L$ and prevent improving the flux noise by shrinking the loop size.
On the other hand, cJJ-based nanoSQUIDs in a simple planar configuration can be fabricated relatively easily from thin film superconductors, e.g., Al, Nb or Pb, through one-step electron-beam (e-beam) or FIB nanopatterning.
Moreover, the use of nanometric-thick films and the smallness of the constriction makes these SQUIDs quite insensitive to in-plane magnetic fields and yields large coupling factors if MNPs are placed close to the constriction [Fig.\ref{Fig:10}(a)].
The small size of cJJs is a key advantage for fabricating nanoSQUIDs with high spin sensitivity.

First thin film Nb dc SQUIDs based on cJJs with linewidths down to 30\,nm, patterned by e-beam lithography, were reported in 1980 \cite{Voss80}.
%
%
Despite their large $L=150\,$pH, miniaturized SQUIDs, with loop size $\sim 1\,\mu{\rm m}^2$, exhibited 
low flux noise $\sim 370\,{\rm n}\Phi_0/\sqrt{\rm Hz}$ at 4.2\,K.
During the 1990s, the use of cJJ nanoSQUIDs for the investigation of small magnetic systems was pioneered by Wernsdorfer {\it et al.}\cite{Wernsdorfer95,Wernsdorfer01,Wernsdorfer09}.
Figure \ref{Fig:10}(b) shows examples of such devices, which were patterned by e-beam lithography from Nb and Al films \cite{Hasselbach00}.
Typical geometric parameters were $1\,\mu{\rm m}^2$ inner loop area, 200\,nm minimum linewidth and   30\,nm film thickness.
%
%
The size of the constrictions ($\sim 30\,$nm wide, $\sim 300\,$nm long) was significantly larger than $\xi$ for Nb.
This lead to a highly non-ideal CPR \cite{Likharev79,Faucher02} and hence non-ideal $I_{\rm c}(\Phi)$ dependence with strongly suppressed $I_c$ modulation depth for Nb cJJ SQUIDs.
Furthermore, $L_{\rm kin}$ of the constrictions can be a few 100\,pH, dominating the overall inductance of the devices \cite{Faucher02}.
Impressively large magnetic fields could be applied parallel to the nanoSQUID loops up to 0.5\,T for Al and 1\,T for Nb.
From the measured critical current noise, the flux noise was calculated as $\sim 40\,\mu\Phi_0/\sqrt{\rm Hz}$ for Al and $\sim 100\,\mu\Phi_0/\sqrt{\rm Hz}$ for Nb \cite{Hasselbach00}.
Due to hysteretic IVCs these nanoSQUIDs were operated in $I_{\rm c}$ readout mode or as threshold detectors (see section \ref{subsubsec:Ic-readout}).
These sensors allowed the vastest realization of true magnetization measurements (section \ref{subsec:Susceptibility}) and were also implemented into probe tips to perform scanning SQUID microscopy \cite{Hasselbach00,Veauvy02}.
%


For similar Nb cJJ-based nanoSQUIDs (30\,nm thick, $\sim 200\,$nm inner loop size, cJJs down to 280\,nm long and 120\,nm wide) switching current distributions were measured from 4.2 down to 2.8\,K \cite{Granata09}.
A detailed analysis of the noise performance for $I_{\rm c}$ readout revealed a flux sensitivity of a few m$\Phi_0$, which was shown to arise from thermally induced $I_{\rm c}$ fluctuations in the nanobridges.
More recently, hysteretic nanoSQUIDs made of Al-Nb-W layers ($2.5\,\mu$m inner loop size; 40\,nm wide, 180\,nm long cJJs) could be operated with oscillating current-bias and lock-in read-out at $T<1.5\,$K \cite{Hazra13}.
In this configuration $I_{\rm c}$ is considerably reduced due to the inverse proximity effect of W on Nb.

Nanometric Nb SQUIDs (50\,nm thick, down to 150\,nm inner hole size) were also fabricated by FIB milling to produce cJJs (80\,nm wide, 150\,nm long) \cite{Troeman07}.
It was observed that Ga implantation depth can reach values of 30\,nm, suppressing the superconducting properties of Nb.
At $T=4.2\,$K, devices with relatively small $I_{\rm c} < 25\,\mu$A showed nonhysteretic IVCs
%
and could be operated in a conventional current-bias mode, yielding $\sqrt{S_{\Phi, \rm w}}\sim 1.5\,\mu\Phi_0/\sqrt{\rm Hz}$.

A possible way to approach the sinusoidal CPR of ideal point contacts is the use of variable thickness nanobridges.
Here, the thicker superconducting banks can serve as phase reservoirs, while the variation in the superconducting order parameter should be confined to the thin part of the bridges \cite{Vijay09}.
cJJ-based nanoSQUIDs were realized by local anodization of ultrathin ($3-6.5\,$nm-thick) Nb films using a voltage-biased atomic force microscope (AFM) tip \cite{Bouchiat01}.
This technique produced constrictions ($30-100\,$nm wide and $200-1000\,$nm long) and variable thickness nanobridges by further reducing the constriction thickness down to few nm (within a $\sim 15\,$nm long section).
The latter exhibited $\Delta I_{\rm c}/I_{\rm c}$ twice as large as the former, indicating an improved CPR.

Vijay {\it et al.} \cite{Vijay10} produced Al nanoSQUIDs based on cJJs (8\,nm thick, 30\,nm wide) with variable length ($l=75-400\,$nm).
The cJJs were either connected to superconducting banks of the same thickness (`'2D devices'') or to much thicker (80\,nm) banks (`'3D devices'').
For 3D devices with $l \le 150\,{\rm nm} \approx 4\xi$, the measured $I_{\rm c}(\Phi)$ curves indicate a CPR which is close to the one for an ideal short metallic weak link.
Both 2D and 3D devices were fully operative up to in-plane magnetic fields of 60\,mT \cite{Antler13}.
Such nanoSQUIDs were operated with dispersive readout (see section \ref{subsubsec:disp-readout}) yielding impressive flux noise values of $30\,{\rm n}\Phi_0/\sqrt{\rm Hz}$ for a 20\,MHz bandwidth \cite{Levenson-Falk13}.

Variable thickness bridges have recently also been realized by connecting suspended Al nanobridges (25\,nm thick, 233\,nm long, 60\,nm wide) to Nb(30\,nm)/Al(25\,nm) bilayer banks to form a nanoSQUID ($2.5\,\mu$m-in-diameter loop)\cite{Hazra14}.
These devices have the advantage of using cJJs from a material (Al) with relatively large $\xi$, while maintaining relatively high $T_{\rm c}$ and critical magnetic field in the superconducting banks forming the SQUID loop.

%
Thermal hysteresis in the IVCs of cJJs can be suppressed by covering the devices with a normal metallic layer, which provides resistive shunting and acts as a heat sink.
cJJ-based nanoSQUIDs from 20\,nm-thick Nb films covered by 25\,nm-thick Au have been patterned by e-beam lithography to realize 200\,nm inner loop size and constriction widths in the range $70-200\,$nm, yielding $L\sim 15\,$pH  \cite{Lam03} .
%
The Au layer prevented hysteresis in the IVCs at temperatures above 1\,K, allowing conventional SQUID readout in the voltage state, yielding $\sqrt{S_{\Phi, \rm w}}\sim 5\,\mu\Phi_0/\sqrt{\rm Hz}$ at 4.2\,K, increasing by about 15\,\% when operating in a magnetic field of 2\,mT \cite{Lam06}.
Field operation up to few 100\,mT was improved by reducing the hole size down to 100\,nm and the largest linewidths down to 250\,nm \cite{Lam11}.
Preliminary experiments were performed on ferritin nanoparticles attached to the cJJs \cite{Vohralik09}.
However, the magnitude of the flux change observed in some cases (up to $440\,\mu\Phi_0$) was larger than the expected one for a ferritin NP located at optimum position (up to $100\,\mu\Phi_0$).

Low-noise nanoSQUIDs from a Nb/amorphous W bilayer (200 and 150\,nm thick, respectively) have been produced by FIB milling \cite{Hao08}.
The SQUID loop (370\,nm inner diameter) was intersected by two nanobridges (65\,nm wide and $60-80\,$nm long)
%
which showed non-hysteretic IVCs at $5-9\,$K.
Readout in the voltage state
%
gave $\sqrt{S_{\Phi, \rm w}} = 200\,{\rm n}\Phi_0/\sqrt{\rm Hz}$ at 6.8\,K.
Recently, the same group extended the operation temperatures down to $<1\,$K by using superconducting Ti films, inversely proximized by Au layers to reduce $T_{\rm c}$ \cite{Blois13}.
These SQUIDs (with 40\,nm wide and 120\,nm long constrictions) exhibited no hysteresis within $60\,{\rm mK} < T < 600\,$mK and had
$\sqrt{S_{\Phi, \rm w}} = 1.1\,\mu\Phi_0/\sqrt{\rm Hz}$.
%
These devices allowed the detection of the magnetic signal produced by a 150\,nm diameter FePt nanobead having $10^7\,\mu_{\rm B}$ at 8\,K in fields up to 10\,mT \cite{Hao11}.
%

As mentioned earlier, cJJ-based nanoSQUIDs can be operated in strong magnetic fields applied in the plane of the loop, which is limited by the upper critical field of the superconductors.
%
%
The use of very thin superconducting layers can increase the effective critical field.
Following this idea, $3-5\,$nm-thick cJJ Nb nanoSQUIDs were fabricated, supporting in-plane fields up to 10\,T.
These sensors proved to be well suited for measuring magnetization curves of microcrystals of Mn$_{12}$ SMMs \cite{Chen10}.
However, their large kinetic inductances lead to large flux noise ($\sim 100\,\mu\Phi_0/\sqrt{\rm Hz}$).
More promising is the use of materials with larger upper critical fields, such as boron-doped diamond  \cite{Mandal11}.
Micrometric SQUIDs based on 100\,nm-wide constrictions in 300\,nm thick films were demonstrated to operate up to impressive fields of 4\,T applied along any direction.
These devices were, however, hysteretic due to heat dissipation.
Flux sensitivity was determined from the critical current uncertainty giving $40\,\mu\Phi_0/\sqrt{\rm Hz}$.

Finally we note that the smallest nanoSQUIDs realized so far, which also include cJJs, are the SQUIDs-on-tip (SOTs) \cite{Finkler10,Vasyukov13}.
These devices will be discussed in more detail in section \ref{sec:SSM}.

\subsubsection{Proximized structures}
\label{subsubsec:proxiJJs}

A normal metal in good contact between superconducting electrodes acquires some of their properties due to the proximity effect, inducing a mini-gap in the density of states of the normal metal.
Andreev pairs can propagate along relatively long distances at low $T$, carrying information on the macroscopic phase of the superconductor.
In the {\it long (short)-junction} regime, when the Thouless energy of the metal is larger (smaller) than the superconducting energy gap, the junction properties will be governed by the normal metal (superconductor).

The first dc SQUID built with {\it long} proximized JJs was based on a CNT intersecting an Al ring \cite{Cleuziou06}.
A gate-modulated supercurrent was demonstrated and flux-induced modulation of the critical current (few nA) was observed at mK temperatures.
The goal was to exploit the small cross section of the CNT ($\sim 1\,{\rm nm}^2$) to provide optimum coupling for molecular nanomagnets attached to it.
An experimental proof-of-principle of such a CNT-based magnetometer is, however, still missing.
A micrometric dc SQUID with graphene proximized junctions (50\,nm long, $4\,\mu$m wide) was also reported \cite{Girit09}.
Flux-induced $I_{\rm c}$ modulation was observed, however, no noise performance of the device was reported.

\begin{figure}[t]
\includegraphics[width=0.7\columnwidth]{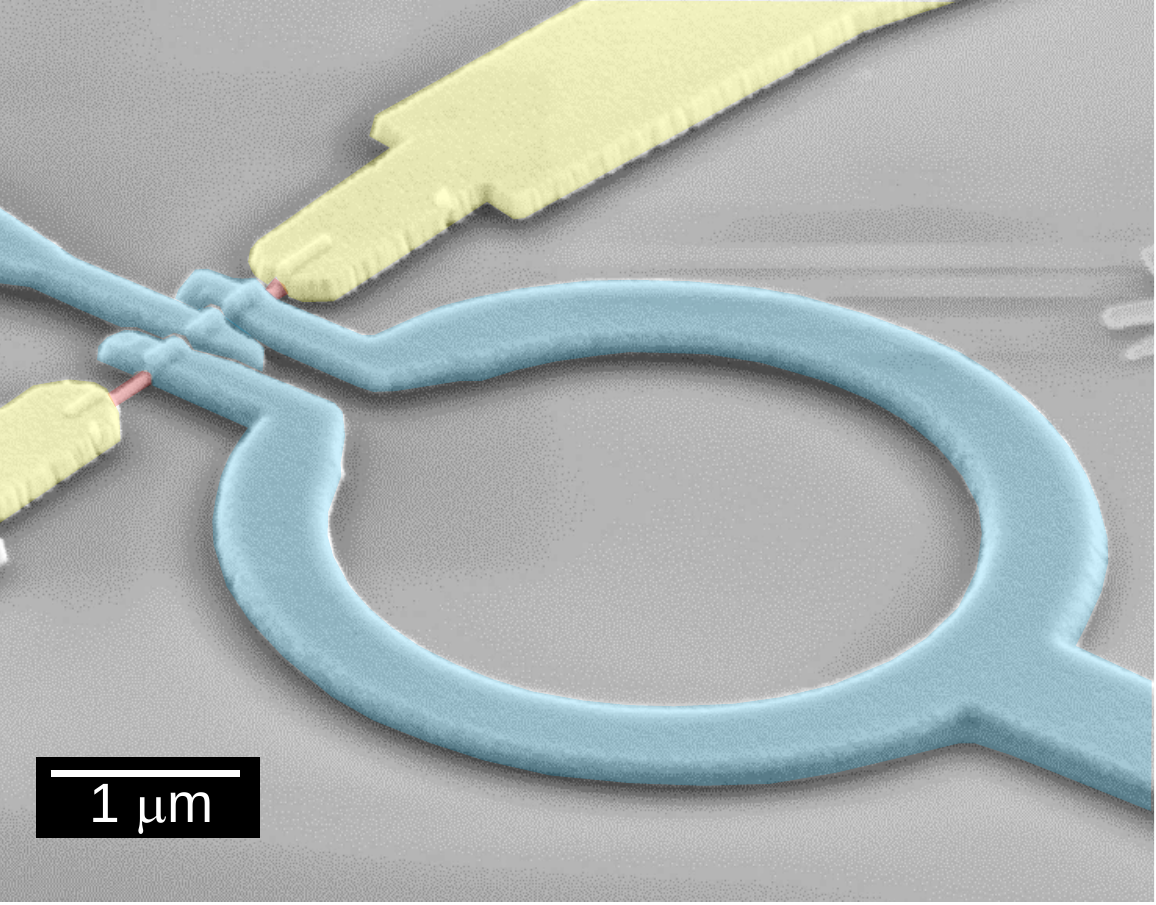}
\caption{SEM image of a SQUID sensor consisting of a proximized highly doped InAs nanowire enclosed within a V ring [after Spathis {\it et al.}\cite{Spathis11}].
SEM image courtesy of F.~Giazotto and S.~D'Ambrosio.}
\label{Fig:11}
\end{figure}

Micrometric dc SQUIDs containing normal metal bridges as weak links have also been reported.
Nb/Au/Nb and Al/Au/Al-based devices showed IVCs with pronounced hysteresis, due to heat dissipated in the normal metal after switching \cite{Angers08}.
SQUIDs with shorter Cu nanowires ($280-370\,$nm long, $60-150\,$nm wide, 20\,nm thick) enclosed in a V ring were non-hysteretic.
%
NanoSQUIDs based on proximized InAs nanowires ($\sim 90\,$nm diameter, 20 or 50\,nm long) were also reported \cite{Spathis11} with JJs in the intermediate length regime [Fig.\ref{Fig:11}].
%


\begin{figure}[b!]
\includegraphics[width=0.7\columnwidth]{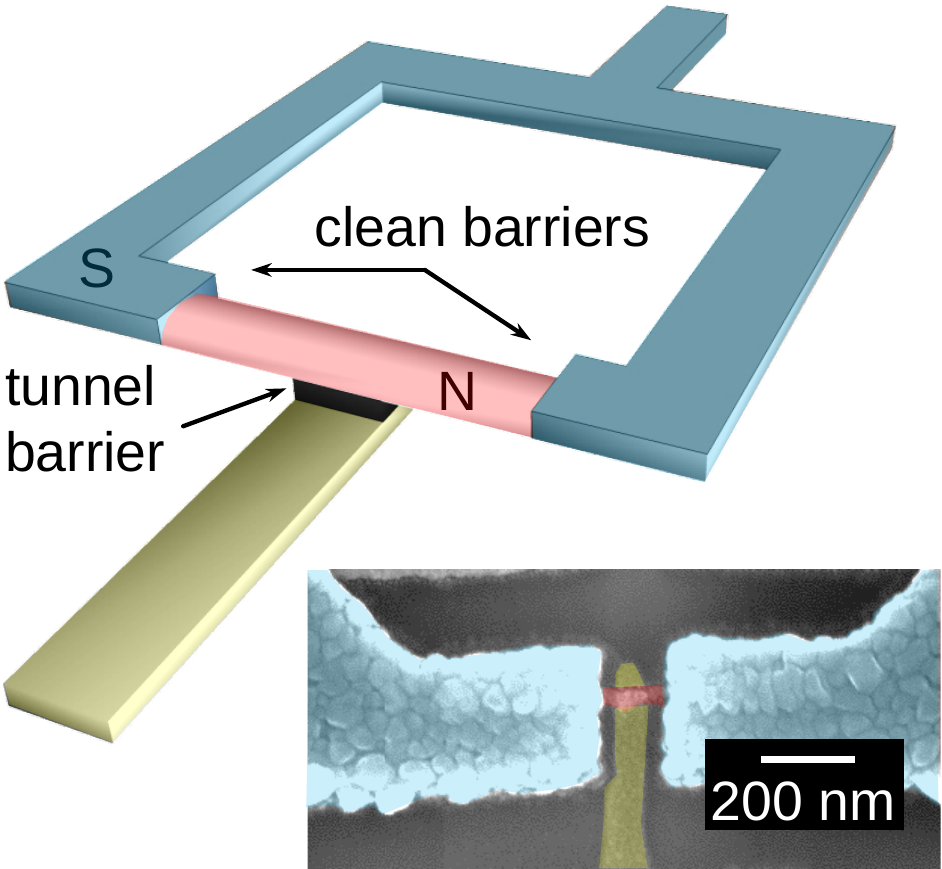}
\caption{Scheme of a SQUIPT.
The inset shows a SEM image of the SQUIPT core; a normal metal probe is tunnel-connected to a proximized Cu island enclosed within an Al ring.
SEM image courtesy of F.~Giazotto and S.~D'Ambrosio.}
\label{Fig:12}
\end{figure}

A different kind of interferometer consists of a superconducting loop interrupted by a normal metal island.
A magnetic field applied to the loop varies the phase difference across the normal metal wire, allowing flux-modulation of the minigap.
This behavior can be probed by an electrode tunnel-coupled to the normal metal island [Fig.\ref{Fig:12}], providing a flux-modulated electric response similar to conventional dc SQUIDs.
This device received the name Superconducting Quantum Interference Proximity Transistor (SQUIPT), for being the magnetic analog to the semiconductor field-effect transistor.
SQUIPTs were pioneered by Giazotto {\it et al.} \cite{Giazotto10} using Al loops and Cu wires ($\sim 1.5\,\mu$m long, $\sim 240\,$nm wide).
These magnetometers were further improved by reducing the length of the normal metal island down to the {\it short-junction} limit, leading to a much larger mini-gap opening.
By choosing proper dimensions of the normal metal island, such sensors do not exhibit any hysteresis down to mK temperatures \cite{Jabdaraghi14,Ronzani14} and can be voltage or current biased, providing impressive values of $V_\Phi$ of a few mV/$\Phi_0$.
SQUIPTs are in their early stage of development\cite{Alidoust13}, still showing a very narrow temperature range of operation limited to sub-kelvin.
On the other hand, they exhibit record low dissipation power of just $\sim 100\,$fW ($I_{\rm c} \sim$pA, $V_{\rm out} \sim 100\,$mV) and should achieve flux noise levels of just a few n$\Phi_0/\sqrt{\rm Hz}$.
The latter has not been determined experimentally yet due to limitations from the voltage noise of the room temperature amplifiers.

\subsection{NanoSQUIDs based on cuprate superconductors}
\label{subsec:high-Tc}

%
High-$T_c$ cuprate superconductors such as YBCO have very small and anisotropic values of $\xi$, reaching $\sim 1\,$nm for the $a-b$ plane and a minute $\sim 0.1\,$nm for the $c$ axis, making the fabrication of cJJs extremely challenging.
Still, the fabrication of YBCO cJJs with $50\,{\rm nm}\times 50\,$nm cross-section and $100-200\,$nm length has been reported recently \cite{Arpaia14}.
These JJs exhibit large $I_{\rm c}$ of a few mA at 300\,mK.
NanoSQUIDs based on this technology were fabricated and preliminary measurements showed low flux noise $\sqrt{S_{\Phi, \rm w}} = 700\,{\rm n}\Phi_0/\sqrt{\rm Hz}$ at 8 K.

Probably the most mature JJs from cuprate superconductors are based on Josephson coupling across grain boundaries (GBs).
Grain boundary junctions (GBJs) can be fabricated, e.g., by epitaxial growth of cuprate superconductors on bicrystal substrates or biepitaxial seed layers \cite{Hilgenkamp02,Tafuri05,Tafuri13}.
Although micrometric SQUIDs based on GBJs have been produced \cite{Koelle99}, the miniaturization of high-quality GBJs is challenging, because of degradation of the material due to oxygen loss during nanopatterning.
NanoSQUIDs made of high-$T_{\rm c}$ GBJs are, on the other hand, very attractive due to their large critical current densities ($\sim 10^5\,{\rm A/cm}^2$ at 4.2\,K) and huge upper critical fields (several tens of T).

\begin{figure}[b]
\includegraphics[width=0.75\columnwidth]{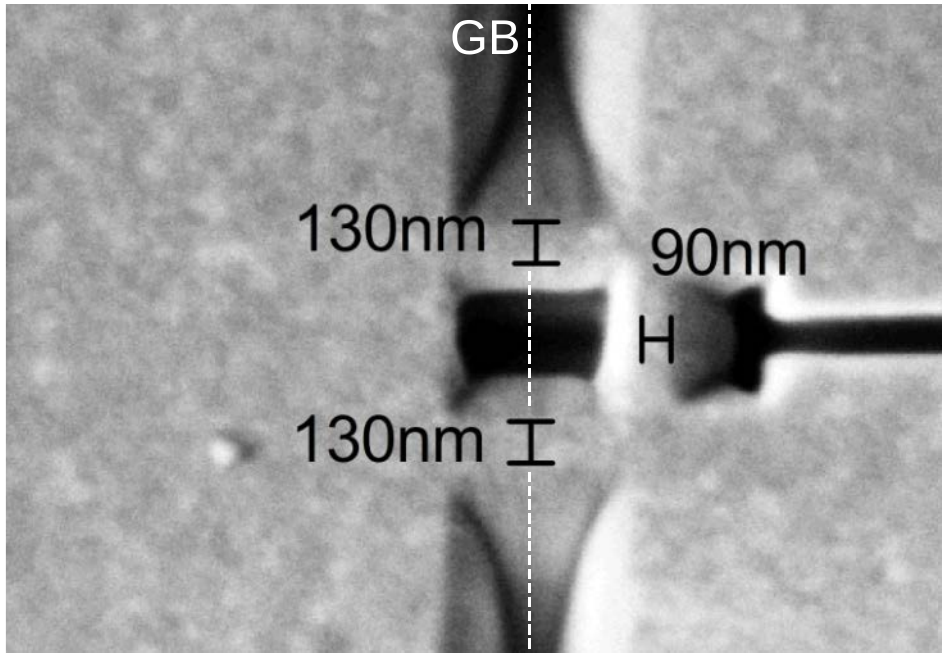}
\caption{SEM image of YBCO nanoSQUID loop ($400 \times 300\,{\rm nm}^2$), intersected by 130\,nm wide GBJs; the GB is indicated by the vertical dashed line.
%
%
The loop contains a 90\,nm wide constriction for flux biasing and optimum coupling.
[after Schwarz {\it et al.}\cite{Schwarz13}]}
\label{Fig:13}
\end{figure}

\begin{figure}[b]
\includegraphics[width=0.9\columnwidth]{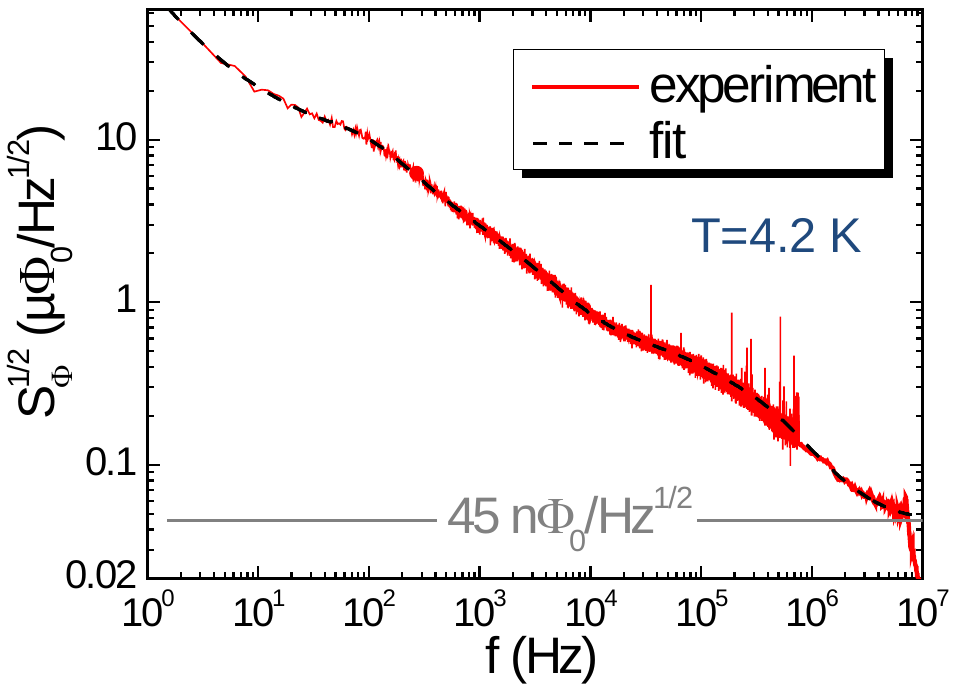}
\caption{(b) Rms flux noise of optimized YBCO nanoSQUID, measured in open-loop mode.
Dashed line is a fit to the measured spectrum; horizontal line indicates fitted white noise.
[after Schwarz {\it et al.}\cite{Schwarz15}]}
\label{Fig:14}
\end{figure}

YBCO GBJ nanoSQUIDs were fabricated by FIB milling \cite{Nagel11,Schwarz13,Schwarz15}.
Devices consist of $50-300\,$nm thick YBCO epitaxially grown on bicrystal SrTiO$_3$ substrates ($24^\circ$ misorientation angle) and covered by typically 60\,nm thick Au serving as resistive shunt and to protect the YBCO during FIB milling.
Typical inner hole size is $200-500\,$nm and GBJs are $100-300\,$nm wide [Fig.~\ref{Fig:13}].
%
%
Devices are non-hysteretic and work from $<1\,$K up to $\sim 80\,$K.
Large magnetic fields can be applied perpendicular to the GBJs in the substrate plane, without severe degradation of the $I_{\rm c}$ modulation for fields up to 3\,T \cite{Schwarz13}.
Via a modulation current $I_{\rm mod}$ through a constriction (down to $\sim 50\,$nm wide) in the loop, the devices can be flux-biased at their optimum working point, without exceeding the critical current, i.e.~the constriction is not acting as a weak link.
The constriction is also the position of optimum coupling of a MNP to the SQUID. 

Numerical simulations based on London equations for variable SQUID geometry provided expressions for $L$ and $\phi_\mu$ [via Eq.~(\ref{eq:phi-mu})] for a magnetic dipole 10\,nm above the constriction, as a function of all relevant geometric parameters.
Together with RCSJ model predictions for $S_{\Phi, \rm w}$ at 4.2\,K, an optimization study for the spin sensitivity has been performed.
An optimum film thickness $d_{\rm opt}=120\,$nm was found (for $\lambda_{\rm L}=250\,$nm).
For smaller $d$, the increasing contribution of $L_{\rm kin}$ to the flux noise dominates over the improvement in coupling.
For optimum $\beta_L \sim 0.5$ and $d=d_{\rm opt}$, the spin sensitivity decreases monotonically with decreasing constriction length $l_{\rm c}$ (which fixes the optimum constriction width $w_{\rm c}$).
For $l_{\rm c}$ and $w_{\rm c}$ of several tens of nm, an optimum spin sensitivity of a few $\mu_{\rm B}/\sqrt{\rm Hz}$ was predicted in the white noise limit \cite{Woelbing14}.

For an optimized device with small inductance $L\sim 4\,$pH ($d=120\,$nm, $l_{\rm c}=190\,$nm, $w_{\rm c}=85\,$nm),
direct readout measurements of the magnetic flux noise at 4.2\,K gave $50\,{\rm n}\Phi_0/\sqrt{\rm Hz}$ at 7\,MHz (close to the intrinsic thermal noise floor), which is amongst the lowest values reported for dc SQUIDs so far [Fig.~\ref{Fig:14}].
With a calculated coupling factor $\phi_\mu=13\,{\rm n}\Phi_0/\mu_{\rm B}$, this device yields a spin sensitivity of $3.7\,\mu_{\rm B}/\sqrt{\rm Hz}$ at 7\,MHz and 4.2\,K \cite{Schwarz15}.
Due to the extremely low white noise level, $1/f$-like excess noise dominates the noise spectrum within the entire bandwidth of the readout electronics.
Bias reversal can only partially eliminate this excess noise, which deserves further investigation.

Finally, an encouraging step towards  the controlled formation and further miniaturization of high-$T_c$ JJs has been made recently by \cite{Cybart15}.
For this purpose a 0.5-nm-diameter He$^+$-beam was used to fabricate $\sim 1\,$nm-narrow ion-irradiated barriers on $4\,\mu$m wide and 30\,nm thick YBCO bridges.
The key point is the smallness of the ion beam diameter, which allows the introduction of point like defects.
By varying the irradiation dose between $10^{14}-10^{18}\,{\rm He}^+/{\rm cm}^2$ the authors showed the successful realization of JJs exhibiting SNS-like or tunnel-like behavior.
This technique has been applied to the fabrication of SQUID devices \cite{Cho15}, but their downsizing to the nanoscale still needs to be realized.

\section{nanoSQUIDs for magnetic particle detection}
\label{sec:MNP-detection}

Originally, nanoSQUIDs were conceived for the investigation of individual MNPs and SMMs.
These systems are of key technological importance with applications ranging from electronics, including hard discs, magnetic random access memories, giant magneto resistance devices, and spin valves, through on-chip adiabatic magnetic coolers, and up to biotechnology applications including enhanced imaging of tissues and organs, virus-detecting magnetic resonance imaging, and cancer therapy (see, e.g., Ref.~\cite{MRSBull13}).
Moreover, magnetic molecules appear as an attractive playground to study quantum phenomena \cite{Bartolome14} and could eventually find application in emerging fields of quantum science such as solid-state quantum information technologies \cite{Leuenberger01} and molecular spintronics \cite{Bogani08}.
%

In this section we will review, as an important application of nanoSQUIDs, the investigation of small magnetic particles.
We will first address challenges and approaches regarding positioning of MNPs close to the SQUIDs and then discuss measurements of magnetization reversal and of ac susceptibility of MNPs.

\subsection{Nanoparticle positioning}
\label{subsec:positioning}

The manipulation and positioning of MNPs close to the nanoSQUIDs is particularly important since the magnetic signal coupled to any form of magnetometer strongly depends on the particle location with respect to the sensor.
Although conceptually very simple, this problem has hampered the realization of true single-particle magnetic measurements so far.
Many strategies have been developed to improve the control on the positioning of MNPs or SMMs on specific areas of nanoSQUID sensors.
%
%

\begin{figure*}[t]
\includegraphics[width=0.7\textwidth]{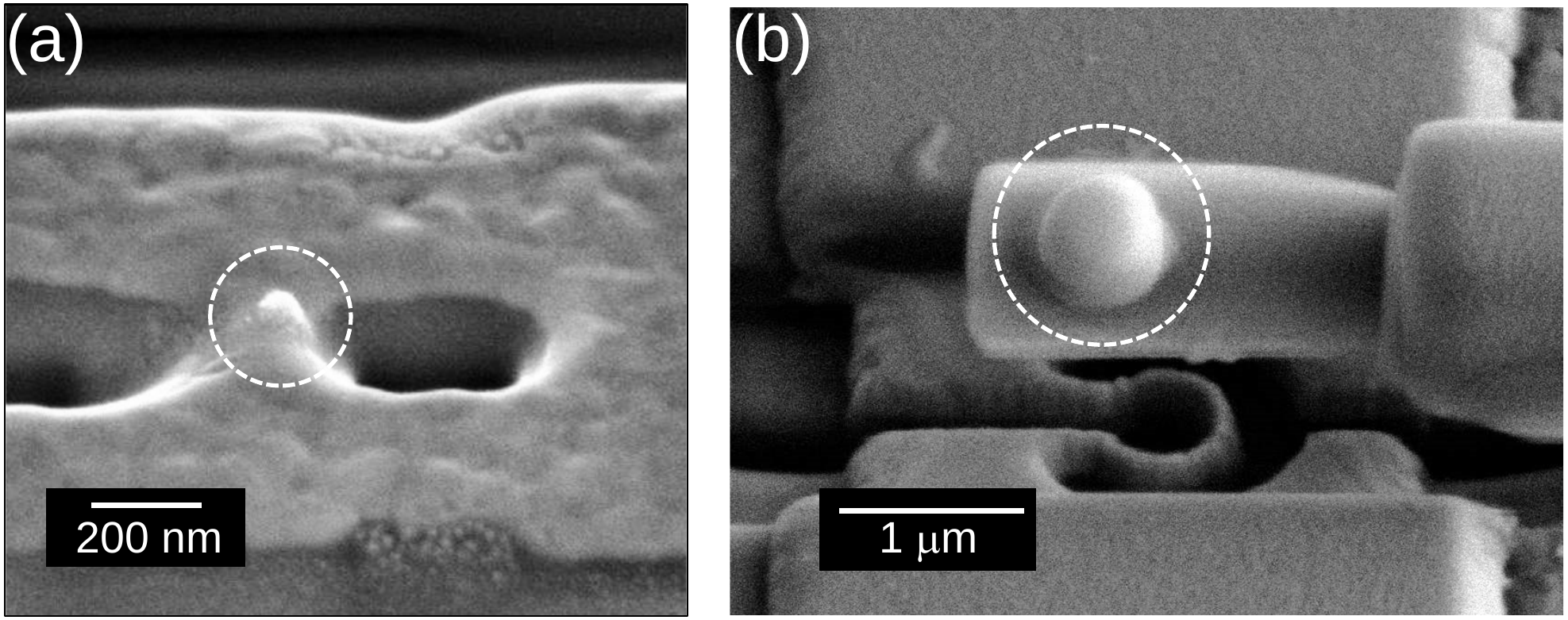}
\hastfill
\begin{minipage}[b]{0.28\textwidth}
\caption{SEM images of (a) Co nanoparticle deposited by FEBID on the constriction of a YBCO nanoSQUID and
(b) nanodot deposited by FIBID on a SiNi cantilever above a Nb nanoloop.
Particles are highlighted by dashed  circles
SEM images courtesy of J.~Ses\'{e}.}
\label{Fig:15}
\end{minipage}
\end{figure*}

\subsubsection{In-situ nanoparticle growth}
\label{subsubsec:grow}

In an early approach, called the drop-casting method, small droplets with suspended MNPs were deposited on a substrate containing many nanoSQUIDs.
After solvent evaporation some of the MNPs happened to occupy positions of maximum coupling.
This method was successfully applied to investigate $15-30\,$nm individual Co MNPs \cite{Wernsdorfer97} .
In a similar approach, 
%
%
MNPs based on Co, Fe or Ni were sputtered using low-energy cluster beam deposition techniques onto substrates containing a large amount of microSQUIDs \cite{Wernsdorfer95a}.
Alternatively, MNP and Nb deposition was realized simultaneously to embed nanometric clusters into the superconducting films, which were subsequently patterned to form nano- or microSQUIDs \cite{Jamet00}.
The drawback of these techniques is the lack of precise control of the MNP positions relative to the SQUIDs, which requires the use and characterization of many tens or even hundreds of SQUIDs.

Improved nanometric control over the particle position can be achieved by nanolithography methods.
This has been used to define Co, Ni, TbFe$_3$ and Co$_{81}$Zr$_9$Mo$_8$Ni$_2$ MNPs with smallest dimension of $100\times 50 \times 8\,{\rm nm}^3$ \cite{Wernsdorfer95}.
Alternatively, focused e-beam induced deposition (FEBID) of high-purity cobalt (from a precursor gas, e.g., Co$_2$(CO)$_8$ \cite{Cordoba10}) allows the definition of much smaller particles (down to $\sim 10\,$nm) and arbitrary shape located at precise positions with nanometric resolution.
%
%
This technique has been successfully applied to the integration of amorphous Co nanodots onto YBCO nanoSQUIDs [Fig.~\ref{Fig:15}(a)]\cite{MartinezPerez17}.

\subsubsection{Scanning probe-based techniques }
\label{subsubsec:scanning-probe}

The use of a scanning probe, e.g., the tip of an AFM can be used for precise manipulation of the position of a MNP.
AFM imaging in non-contact mode is first used to locate MNPs dispersed over a surface
Then, using contact mode, the tip is used to literally `'push'' the MNP to the desired position \cite{Martin98,Pakes04}.
This technique was applied to improve the coupling between a nanoSQUID and Fe$_3$O$_4$ NPs (15\,nm diameter) deposited via the drop-casting method \cite{Wernsdorfer09}.
Micro- and nanomanipulators installed inside SEMs have also been used for this purpose.
For instance, a sharpened carbon fiber mounted on a micromanipulator in a SEM has been used to pick up a $\sim 0.15\,\mu$m diameter single FePt particle an deposit it onto a nanoSQUID \cite{Hao11}.

Alternatively, larger carriers that are more easily visible and manipulated can be used to manipulate the position of MNPs.
For example, microscopic SiNi cantilevers containing the MNP of interest can be moved using a micromanipulator 
\cite{Gella15} [Fig.~\ref{Fig:15}(b)]. 
In particular, CNTs appear as promising tools for this purpose.
SMMs have indeed been successfully grafted over or encapsulated inside CNTs, which were later used to infer their magnetic properties \cite{Ganzhorn13}.
Similarly, an Fe nanowire encapsulated in a CNT has been mounted by micromanipulators on top of YBCO nanoSQUIDs for magnetization reversal measurements (see section \ref{subsec:Magnetization}) \cite{Schwarz15}.

\begin{figure*}[t]
\includegraphics[width=0.7\textwidth]{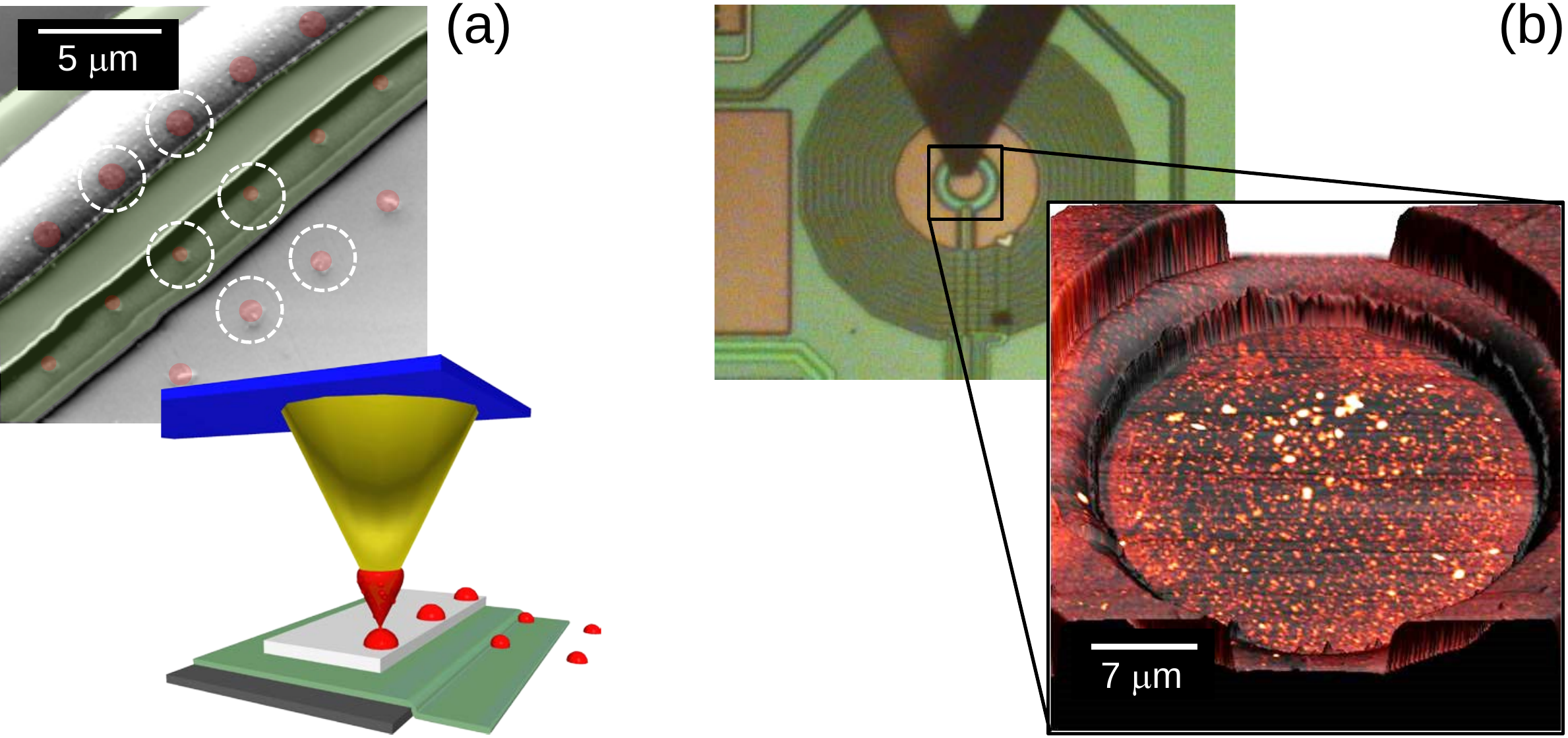}
\caption{(a) Ferritin nanodots (dashed circles) deposited by DPN on top of the pickup coil of a SQUID-based microsusceptometer.
Each dot contains $10^4$ proteins approximately arranged as a monolayer.
Scheme of the DPN nanopatterning technique; a conventional AFM probe delivers dot-like features containing monolayer arrangements of ferritin over the surface [after Mart\'{i}nez-P\'{e}rez {\it  et al.}\cite{Martinez-Perez11a}].
(b) Optical microscope image taken during the DPN patterning process showing the AFM probe over a microsusceptometer's pickup coil.
The blow-up shows an AFM image of the resulting sample containing 5 molecular layers of Dy$_2$ SMMs.
Images courtesy of F.~Luis.}
\label{Fig:16}
\end{figure*}

Another promising approach is dip pen nanolithography (DPN).
%
%
Here, an AFM tip is first coated with a solution containg MNPs and then brought into contact with a surface at the desired location.
Capillarity transport of the MNPs from the tip to the surface via a water meniscus enables the successful deposition of small collections of molecules in submicrometer dimensions \cite{Piner99}.
Bellido {\it et al.} \cite{Bellido10} showed that this technique can be applied to the deposition of dot-like features containing monolayer arrangements of ferritin-based molecules onto microSQUID sensors [Fig.~\ref{Fig:16}(a)] for magnetic susceptibility measurements \cite{Martinez-Perez11a} (section \ref{subsec:Susceptibility}).
%
%
%
The number of MNPs deposited per dot can be controlled (via the concentration of the ferritin solution and dot size) from several hundred of proteins down to individual ones \cite{Bellido10}.
Recently, DPN has also been applied to the deposition of dot-like features containing just $3-5$ molecular layers of Mn$_{12}$ and Dy$_2$ SMMs onto the active areas of microSQUID-based susceptometers, enabling the detection of their magnetic susceptibility \cite{Bellido13,Jenkins15} [Fig.~\ref{Fig:16}(b)].
%

Recently, individual magnetic nanotubes, attached to an ultrasoft cantilever were brought in close vicinity to a nanoSQUID at low $T$ \cite{Buchter13,Nagel13,Buchter15}.
This technique allowed the authors to 
investigate magnetization reversal of the nanotubes by combining torque and SQUID magnetometry (see section \ref{subsec:Magnetization}).

We note that scanning SQUID microscopy could also be applied to the study of MNPs deposited randomly on surfaces \cite{Kirtley09}.
This would provide an elegant way of locating magnetic systems close enough to the sensor and would also enable in-situ reference measurements.
However, their use for the investigation of magnetic molecules or nanoparticles arranged on surfaces is still awaiting.
%

\subsubsection{Techniques based on chemical functionalization}
\label{subsubsec:chem-funct}

The above mentioned techniques can be further improved by chemically functionalizing the sensor's surface or the MNPs or both of them \cite{Bellido12}.
This usually provides high quality monolayers or even individual magnetic molecules at specific positions.
For instance, Mn$_{12}$ SMMs could be successfully grafted on Au, the preferred substrate for chemical binding, by introducing thiol groups in the clusters \cite{Cornia03}.
In a further step, such Mn$_{12}$ molecules could be individually isolated by a combination of molecule and Au substrate functionalization \cite{Coronado05}.

This technique has also been applied to the deposition of ferritin-based MNPs onto Au-shunted nanoSQUIDs \cite{Lam08}.
For this purpose, a $200 \times 200\,{\rm nm}^2$ window was opened through e-beam lithography onto a PMMA layer deposited on top of the nanoSQUID.
This window was then covered with organic linkers that were later used to attach the ferritin MNPs.
The success of this process was finally determined by AFM, showing evidence that few proteins were attached.
%

\subsection{Magnetization measurements}
\label{subsec:Magnetization}

NanoSQUIDs can be applied to study the reversal of magnetization $M$ of MNPs placed nearby.
For this purpose an external magnetic field $B_{\rm ext}$ is swept while recording changes in the magnetic moment $\mu$ of the sample coupled as a change of magnetic flux to the SQUID [Fig.~\ref{Fig:4}(b)].
Usually, $M(B_{\rm ext})$ is hysteretic, due to an energy barrier created by magnetic anisotropy.
Such hysteresis loops reveal information on the reversal mechanisms, e.g.~domain wall nucleation and propagation or the formation of topological magnetic states like vortices, coherent rotation, or quantum tunneling of magnetization.
Depending on the particle's anisotropy, this requires the application of relatively large $B_{\rm ext}$, a difficult task when dealing with superconducting materials.
Measurements are usually done by careful alignment of $B_{\rm ext}$ with respect to the nanoSQUID, to minimize the magnetic flux coupled to the loop and the JJs by $B_{\rm ext}$ directly.
The maximum $B_{\rm ext}$ will be limited by the upper critical field of the superconducting material, e.g.~$\sim 1\,$T for Nb films, unless ultrathin films are used, which however increases significantly $L_{\rm k}$ and hence the flux noise (see section \ref{subsubsec:cJJs}).

\begin{figure*}[t]
\includegraphics[width=0.6\textwidth]{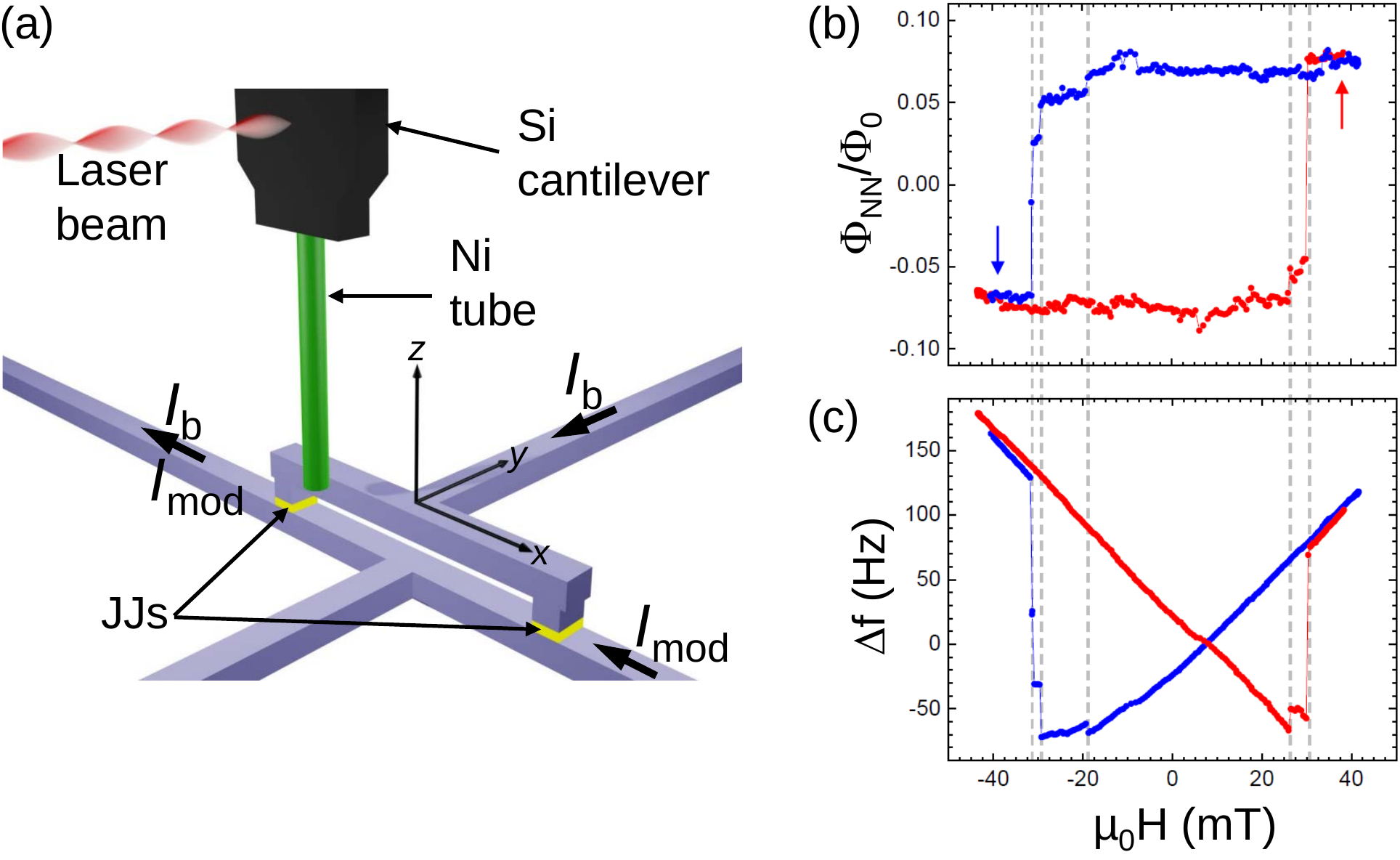}
\hastfill
\begin{minipage}[b]{0.38\textwidth}
\caption{(a) Sketch of combined torque and nanoSQUID magnetometry on a Ni nanotube. 
%
(b,c) Simultaneously measured hysteresis loops (b) $\Phi_{\rm NN}(H)$, (c) $\Delta f(H)$. 
Arrows indicate $H$ sweep direction.
Dashed lines indicate discontinuities 
appearing in both $\Phi_{\rm NN}(H)$ and $\Delta f(H)$.
[after \cite{Nagel13} and \cite{Buchter13}]}
\label{Fig:17}
\end{minipage}
\end{figure*}

The vastest amount of dc magnetization studies performed on individual MNPs was provided by the pioneering work of Wernsdorfer and co-workers.
They were able to measure magnetization curves of a number of MNPs made of Ni, Co, TbFe$_3$ and Co$_{81}$Zr$_9$Mo$_8$Ni$_2$ with sizes down to $100 \times 50 \times 8\,{\rm nm}^2$.
Furthermore, they succeeded in measuring the dc magnetization of the smallest MNPs ever detected to date.
These are 3\,nm diameter crystalline Co MNPs ($10^3\,\mu_{\rm B}$ each) directly embedded into the Nb film forming the nanoSQUID \cite{Jamet01}.
The detected magnetization switching process was attributed to an individual MNP located precisely at the cJJ, where the coupling factor is maximized.
These studies also enabled the determination of the $2^{\rm nd}$ and $4^{\rm th}$ order anisotropy terms in the magnetic anisotropy of the Co MNPs.
Additionally, many exciting phenomena were studied with this technique.
These include, e.g., the observation of Stoner-Wohlfarth and N\'{e}el-Brown type of thermally assisted magnetization reversal in individual Co clusters (25\,nm, $10^6\,\mu_{\rm B}$) \cite{Wernsdorfer97} or the observation of macroscopic quantum tunneling of magnetization in BaFeCoTiO single particles ($10-20\,$nm, $10^5\,\mu_{\rm B}$) \cite{Wernsdorfer97a}.
Magnetization reversal triggered by rf field pulses on a 20\,nm diameter Co NP was also reported \cite{Thirion03} and, recently, the effects of the antiferromagnetic-ferromagnetic exchange bias between a Co nanocluster and a CoO layer were revealed \cite{LeRoy11}.
Micrometric SMM crystals were also investigated with an array containing four microSQUIDs \cite{Wernsdorfer96}.
These experiments allowed observing the modulation of the small ($10^{-7}\,$K) tunnel splitting in Fe$_8$ molecular clusters under the application of a transverse magnetic field \cite{Wernsdorfer99}.

Magnetization reversal mechanisms in single Ni and permalloy nanotubes were investigated using Nb/HfTi/Nb-based nanoSQUIDs \cite{Buchter13,Nagel13,Buchter15}.
Experiments were performed at 4.2\,K with $B_{\rm ext}=\mu_0H$ applied along the nanotube axis ($z$-axis), with the SQUID loop in the $x$-$z$ plane.
The nanoSQUID was mounted on an $x$-$y$-$z$ stage below the bottom end of the nanotube which is affixed to an ultrasoft Si cantilever [Fig.\ref{Fig:17}(a)].
The nanotube was positioned to maximize the flux $\Phi_{\rm NN}$ coupled to the nanoSQUID.
While recording the SQUID output operated in FLL, simultaneously the magnetic torque exerted on the nanotube was detected, by recording the frequency shift $\Delta f$ on the cantilever resonance frequency as a function of $H$.
Measurements on a Ni nanotube showed discontinuities at the same values of $H$ that were ascribed to switching of the magnetization along the nanotube [Fig.~\ref{Fig:17}(b)].
These experiments provided, on the one hand, the magnetic field stray produced by the nanotube's end and, on the other, the volume magnetization, giving evidence for the formation of a magnetic vortex-like configuration in the nanotube.
Measurements on an individual permalloy nanotube evidenced the nucleation of magnetic vortices at the nanotube's end before propagating through its whole length, leading to the complete switching of the magnetization.
Furthermore, it has been shown that a thin exchange-coupled antiferromagnetic native-oxide layer on the nanotube modifies the magnetization reversal process at low temperatures \cite{Buchter15}.

\begin{figure}[b]
\includegraphics[width=0.7\columnwidth]{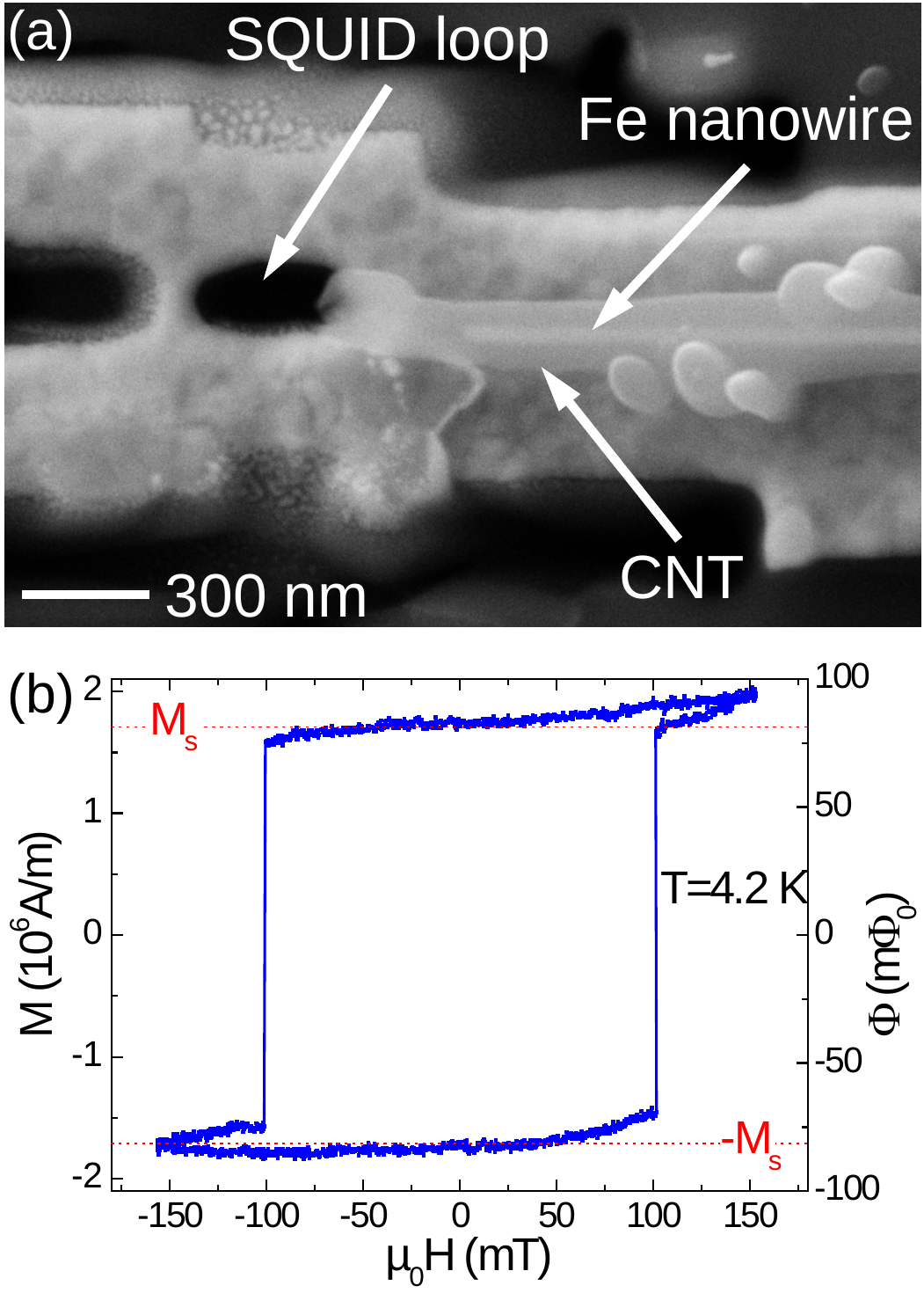}
\hastfill
\begin{minipage}[b]{0.28\columnwidth}
\caption{\\(a) SEM image of Fe nanowire encapsulated in a CNT on top of a YBCO nanoSQUID.
(b) Hysteresis loop $\Phi (H)$ of the Fe nanowire, detected by the SQUID.
Left axis corresponds to magnetization signal $M$; the literature value for the saturation magnetization $M_{\rm s}=1710\,$kA/m of Fe is indicated as dashed lines.
[after Schwarz {\it et al.}\cite{Schwarz15}]}
\label{Fig:18}
\end{minipage}
\end{figure}

YBCO nanoSQUIDs were used for the investigation of magnetization reversal in a Fe nanowire grown inside a CNT attached on top of the SQUID \cite{Schwarz15}  [Fig.\ref{Fig:18}(a)].
Magnetization measurements were performed at 4.2\,K in FLL by continuously sweeping $H$ in the plane of the SQUID loop, along the Fe wire axis.
Rectangular shaped hysteresis loops [Fig.\ref{Fig:18}(b)] indicate a single domain state for the nanowire.
The magnitude of the switching field suggests that magnetization reversal takes place non-uniformly, e.g., by curling.
These results agree very well with previous measurements on an individual nanowire using a micro-Hall bar \cite{Lipert10}, albeit with a significantly improved signal-to-noise ratio.
Similarly, YBCO nanoSQUIDs were used to detect the magnetization reversal of individual Co MNPs with magnetic moments $(1 - 30)\times10^6$ $\mu_{\rm B}$ at different temperatures ranging from 300 mK up to 80 K.
These studies allowed the identification of two different reversal mechanisms which depend on the dimensions and shape of the Co particles. 
The different reversal mechanisms are linked to the stabilization two different magnetic states, i.e., the (quasi) single-domain and the vortex state \cite{MartinezPerez17}.

\subsection{Susceptibility measurements}
\label{subsec:Susceptibility}

Even more demanding, nanoSQUIDs can also be used to quantify the response of a MNP to an oscillating magnetic field $B_{\rm ac}=B_0\cos(\omega t)$, i.e., its frequency-dependent magnetic susceptibility $\chi_{\rm ac}=\chi_{\rm re} + i\chi_{\rm im}$, where $\chi_{\rm re}$ is the part going in-phase with $B_{\rm ac}$ and $\chi_{\rm im}$ is the out-of-phase part.
These quantities bear much information on the dynamic behavior of spins and the relaxation processes to thermal equilibrium, the interaction between spins, and the ensuing magnetic phase transitions.
These measurements can be performed using SQUID-based susceptometers, usually in a gradiometric design to be insensitive to homogeneous external magnetic fields, but sensitive to the imbalance produced by a MNP located in one of the coils [Fig.~\ref{Fig:4}(c,d)].
$\chi_{\rm re}$ and $\chi_{\rm im}$ are directly accessible by applying a homogeneous $B_{\rm ac}$ via on-chip excitation coils and lock-in detecting the nanoSQUID output.
Alternatively, $\sqrt{S_\Phi}$ can be measured, as it is directly related to $\chi_{\rm im}$ through the fluctuation-dissipation theorem \cite{Reim86}.
The detection of $\chi_{\rm ac}$ demands high sensitivity, as the net oscillating polarization induced in the sample is, by far, smaller than the total saturation magnetization.
At best, broad-band frequency measurements must be performed which also provide an easy way to filter out the $1/f$ noise of the SQUIDs, therefore improving the effective sensitivity of the sensor.
Frequencies are usually restricted to $\sim 1\,$MHz, mainly limited by the room temperature amplifiers and the FLL circuit.

One of the most controversial observations of quantum coherence in nanoscopic magnets was realized using the SQUID-based microsusceptometer developed by Ketchen {\it et al.} \cite{Ketchen84}.
%
This device allowed the detection of the magnetic susceptibility of small spin populations of natural horse-spleen ferritin \cite{Awschalom92}.
For a sample with just $4 \times 10^4$ proteins ($\sim 200\,\mu_{\rm B}$/protein), a resonance peak in both the out-of-phase component of $\chi_{\rm ac}$ and $\sqrt{S_\Phi}$ has been observed and was attributed to the zero-field splitting energy \cite{Awschalom92,Awschalom92a}.
This is the energy separating the two non-degenerated low-energy quantum states, i.e., the (anti-)symmetric combination of the classical states corresponding to magnetization pointing (down) up.
This interpretation and the magnitude of this zero-field splitting (900\,kHz) is still an object of debate.

\begin{figure}[t]
\includegraphics[width=\columnwidth]{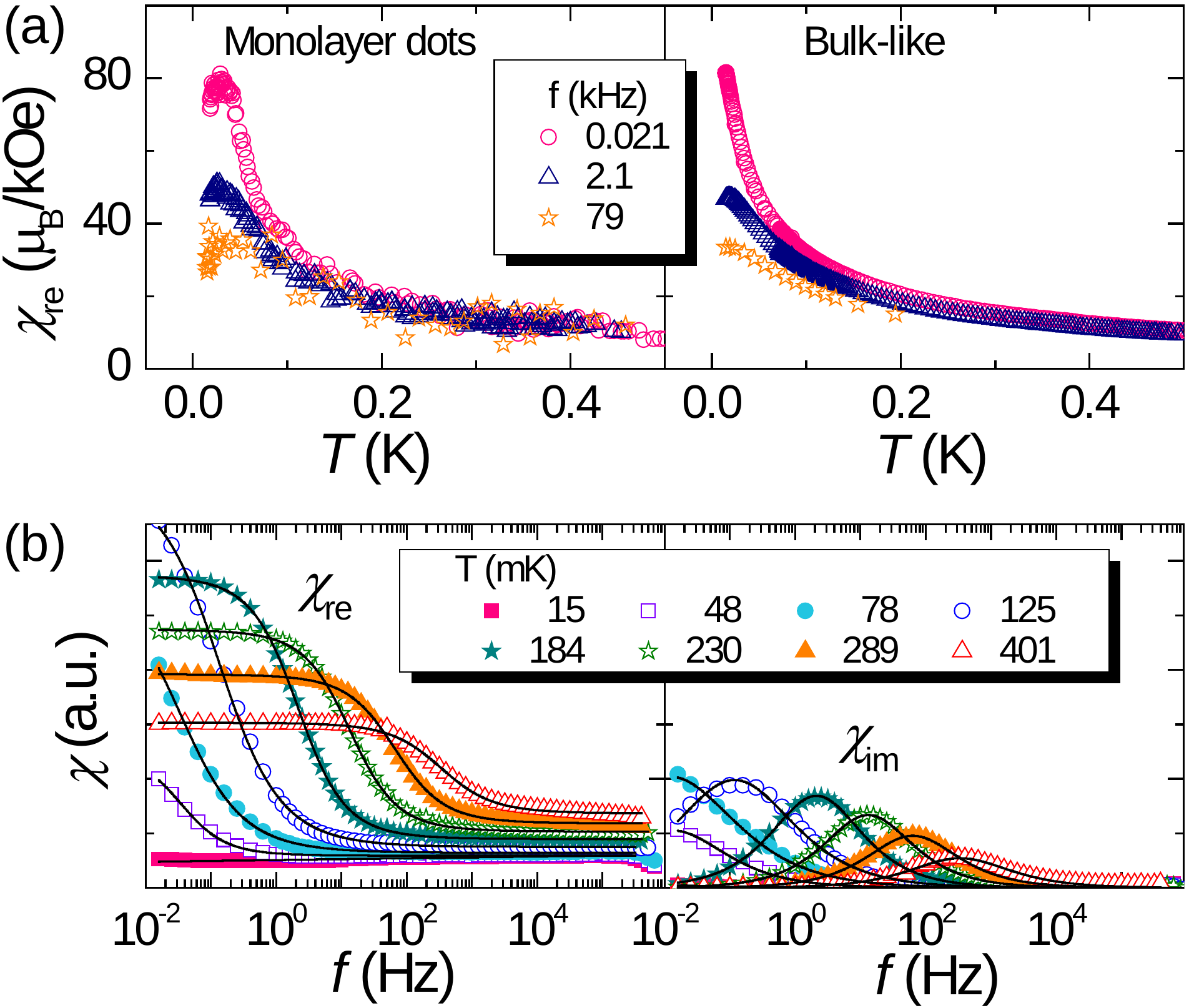}
%
\caption{Magnetic susceptibility $\chi$ measured with SQUID-based microsusceptometers.
(a) Ferritin monolayer dots and bulk sample: $\chi_{\rm re}(T)$ obtained at three different frequencies.
%
%
The superparamagnetic blocking of the susceptibility is visible below 50\,mK in both cases
[after Mart\'{i}nez-P\'{e}rez {\it et al.}\cite{Martinez-Perez11a}].
(b) HoW$_{10}$ SMM crystal: $\chi_{\rm re}(f)$ (left) and $\chi_{\rm im}(f)$ (right) measured at different $T$.}
\label{Fig:19}
%
\end{figure}

MNPs artificially grown inside ferritin were also studied using a SQUID-based microsusceptometer \cite{Martinez-Perez11a}.
The magnetic core with diameter of just a few nm was composed of antiferromagnetic CoO leading to a tiny magnetic moment of $\sim 10\,\mu_{\rm B}$ per protein.
Monolayer arrangements of ferritin MNPs (total amount $\sim 10^7$ proteins) were deposited by DPN directly onto the SQUID, maximizing the coupling between the samples and the sensor's pickup coils \cite{Bellido10} (see section \ref{subsubsec:scanning-probe}).
Using $B_{\rm ac} \sim 0.1\,$mT, these experiments showed that ferritin-based MNPs arranged on surfaces retain their properties, still exhibiting superparamagnetic blocking of the magnetic susceptibility [Fig.~\ref{Fig:19}(a)].
Furthermore, these results allowed to determine experimentally the spin sensitivity.
This was done by determining the coupling, i.e., the measured flux signal coupled to the microsusceptometer divided by the total magnetic moment of the particle, which was located at an optimum position on top of the field coil or close to the edge of the pickup-loop.
Together with the measured flux noise of the SQUID, this yielded $S_\mu^{1/2}\sim 300\,\mu_{\rm B}/\sqrt{\rm Hz}$.
Additionally, a large amount of measurements on SMM micron-sized crystals or powder at very low $T$ were reported [Fig.~\ref{Fig:19}(b)].
The large bandwidth of these susceptometers (1\,mHz--1\,MHz) enabled, e.g., the investigation of the relationship between quantum tunneling and spin-phonon interaction and to point out novel and reliable molecular candidates for quantum computing and low-temperature magnetic refrigerants (e.g., Refs.~\cite{Luis11,Martinez-Perez11,Martinez-Perez12,Martinez-Perez12a}).

Microsusceptometers were also used to detect the ac magnetic susceptibility of just $\sim 9\times 10^7$ Mn$_{12}$ SMMs arranged as dot-like features containing 3--5 molecular layers \cite{Bellido13}.
Measurements showed an evident decrease of the magnetic relaxation time compared to that observed in crystalline Mn$_{12}$.
This phenomenon was attributed to structural modifications of the surface-arranged molecules leading to an effective decrease of their activation energy.
These sensors have been also applied to the investigation of quantum spin dynamics of Fe$_4$ SMMs grafted onto graphene flakes \cite{Cervetti16}.

\section{nanoSQUIDs for scanning SQUID microscopy}
\label{sec:SSM}

In scanning SQUID microscopy (SSM) the high sensitivity of SQUIDs  to magnetic flux is combined with high spatial resolution by scanning a sample under investigation relative to a miniaturized  SQUID sensor, or vice versa.
A variety of SSM systems has been developed in the 1990s and refined since then.
Those were based on both, metallic low-$T_c$ and high-$T_c$ cuprate superconductors, although the majority of work focused on the low-$T_c$ devices.
For a review on the developments of SSM in the 1990s see Ref.~\cite{Kirtley02}.

Obviously, miniaturized SQUID structures can significantly improve the spatial resolution and sensitivity to local magnetic field sources.
A key issue is the requirement to approach the surface of the samples under investigation to a distance which is of the order of or even smaller than the SQUID size or pickup loop, in order to gain in spatial resolution by shrinking the lateral dimensions of the structures.
Several strategies for improving the spatial resolution in SSM have been followed, which can be divided into three approaches.
The two conventional approaches, developed in the 1990s use SQUID structures on planar substrates.
One is based on the sensing of local fields by a miniaturized pickup loop, coupled to a SQUID sensor; the other is based on using miniaturized SQUID loops to which local magnetic signals are coupled directly (section \ref{subsec:SSM-planar}).
A very recently developed third approach uses the SQUID-on-tip (SOT), i.e.~a SQUID deposited directly on top of a nanotip (section \ref{subsec:SOT}).

\subsection{SQUID microscopes using devices on planar substrates}
\label{subsec:SSM-planar}

SQUID microscopes developed at IMB research by Kirtley {\it et al.} \cite{Kirtley95} are based on Nb/Al-AlO$_x$/Nb technology.
The sensors are based on a single SQUID loop with an integrated pickup loop \cite{Ketchen89}.
The pickup loops have diameters down to $\sim 4\,\mu$m and are connected via well shielded superconducting thin film leads to the SQUID loop at typically $\sim 1\,$mm distance on the same chip \cite{Ketchen95}.
This technology has also been used to realize a miniature vector magnetometer for SSM by using three SQUIDs with orthogonal pickup loops on a single chip \cite{Ketchen97}.
As a key advantage, the IBM designs are based on the very mature Nb multilayer SIS technology, including patterning by photolithography, that allows e.g.~using the HYPRES \footnote{www.hypres.com} process for sensor fabrication.
Moreover, this allows to integrate field coils around the pickup loop for susceptibility measurements and modulation coils inductively coupled to the SQUID loop for separate flux modulation of the SQUID, i.e.~without disturbing the signals to be detected by the pickup loop.
The Si substrate is polished to form a corner, typically at a distance $d_{\rm corner}$ of a few tens of $\mu$m away from the center of the pickup loop.
SQUID microscopes based on such sensors use a mechanical lever for scanning.
The SQUID chip is mounted on a cantilever with a small inclination angle $\alpha$ to the plane of the sample.
The vertical pickup-loop to sample distance is then given by $d_{\rm corner}\sin{\alpha}$ \cite{Kirtley95}.
If the SQUID is well thermally linked to the liquid He bath for operation at 4.2\,K, the sample mounted in vacuum can be heated to above $\sim 100\,$K \cite{Kirtley99b}.

The most important application of the IBM microscope was the pioneering work on the order parameter symmetry of cuprate superconductors.
Just to mention a few examples, this includes key experiments for providing clear evidence of $d_{x^2-y^2}$-wave pairing in the cuprates by imaging fractional vortices along YBCO GBJs \cite{Mannhart96}, the formation of half-integer flux quanta in cuprate tricrystals \cite{Tsuei94} and in Nb/cuprate hybrid Josephson junctions, forming zigzag-type JJs  or huge arrays of $\pi$-rings \cite{Hilgenkamp03}.
For more applications, see the review \cite{Kirtley10}.
%


Very similar devices, also based on the Nb multilayer technology, have been developed and used for SSM by the Stanford group of Moler and co-workers \cite{Gardner01,Huber08}.
Based on the original microsusceptometer design of Ketchen {\it et al.} \cite{Ketchen89}, these devices contain two oppositely wound pickup coils, to cancel homogeneous applied fields.
Sensors with $\sim 4\,\mu$m pickup-loop diameter achieved $\sqrt{S_\Phi}=0.8\,\mu\Phi_0/\sqrt{\rm Hz}$ at 4\,K and $0.25\,\mu\Phi_0/\sqrt{\rm Hz}$ below 0.5\,K \cite{Huber08,Koshnick08}.
The sensor's substrate was cut by polishing, leading to $d_{\rm corner}\sim 25\,\mu$.
A capacitive approach control was used to monitor the probe-to-sample distance.
These microsusceptometers were largely improved by using a terraced cantilever obtained through a multilayer lithography process.
In this way the pickup loop stands above the rest of the structure lying at just 300\,nm above the sample surface.
Additionally the pickup loop diameters were reduced down to 600\,nm using focused ion beam (FIB) milling \cite{Koshnick08}.
Based on these SQUID sensors, the Stanford group has developed a SQUID microscope operating at temperatures down to 20\,mK in a dilution refrigerator\cite{Bjoernsson01}.

The SSM system of the Stanford group has been very successfully applied to a variety of interesting systems.
Just to give a few examples, this includes the study of edge currents in topological insulators \cite{Nowack13}, surface magnetic states \cite{Bert11} and twin walls \cite{Kalisky13} at the LaAlO$_3$/SrTiO$_3$ interface, or unpaired spins in metals \cite{Bluhm09}.

As an alternative approach, the group of Hasselbach and co-workers at Institut N\'{e}el, Grenoble developed an SSM based on miniaturized Nb and Al SQUIDs loops with constriction JJs \cite{Hasselbach00}, very similar to the ones of the Wernsdorfer group \cite{Wernsdorfer09}.
This approach allows for a relatively simple single layer fabrication process with prospects of strong miniaturization.
To achieve at the same time small probe-to-sample distances, the sensor's substrate was cut using a dicing machine and a mesa was defined by means of reactive ion etching so that the distance between the SQUID and apex of the mesa (`tip') was only $2-3\,\mu$m.
With an inclination angle $\alpha\sim5^\circ$, this gives a smallest vertical distance to a sample surface of $\sim 0.26\,\mu$m.
The SSM setup is combined with force microscopy, based on the use of a mechanically excited quartz tuning fork and operates in a dilution refrigerator, achieving minimum SQUID and sample temperatures of 0.45\,K \cite{Veauvy02}. 
Very recently, in a modified setup, a SQUID-to-sample distance of 420\,nm has been demonstrated, in a setup with 40\,mK base temperature \cite{Hykel14}.

The SSM system of the Grenoble group has been applied to the investigation of basic properties of superconductors.
This includes, e.g., studies on the direct observation of the localized superconducting state around holes in perforated Al films \cite{Veauvy04} or on the Meissner-Ochsenfeld effect and absence of the Meissner state in the ferromagnetic superconductor UCoGe \cite{Paulsen12}.

\subsection{SQUID-on-tip (SOT) microscope}
\label{subsec:SOT}

\begin{figure}[t]
\includegraphics[width=\columnwidth]{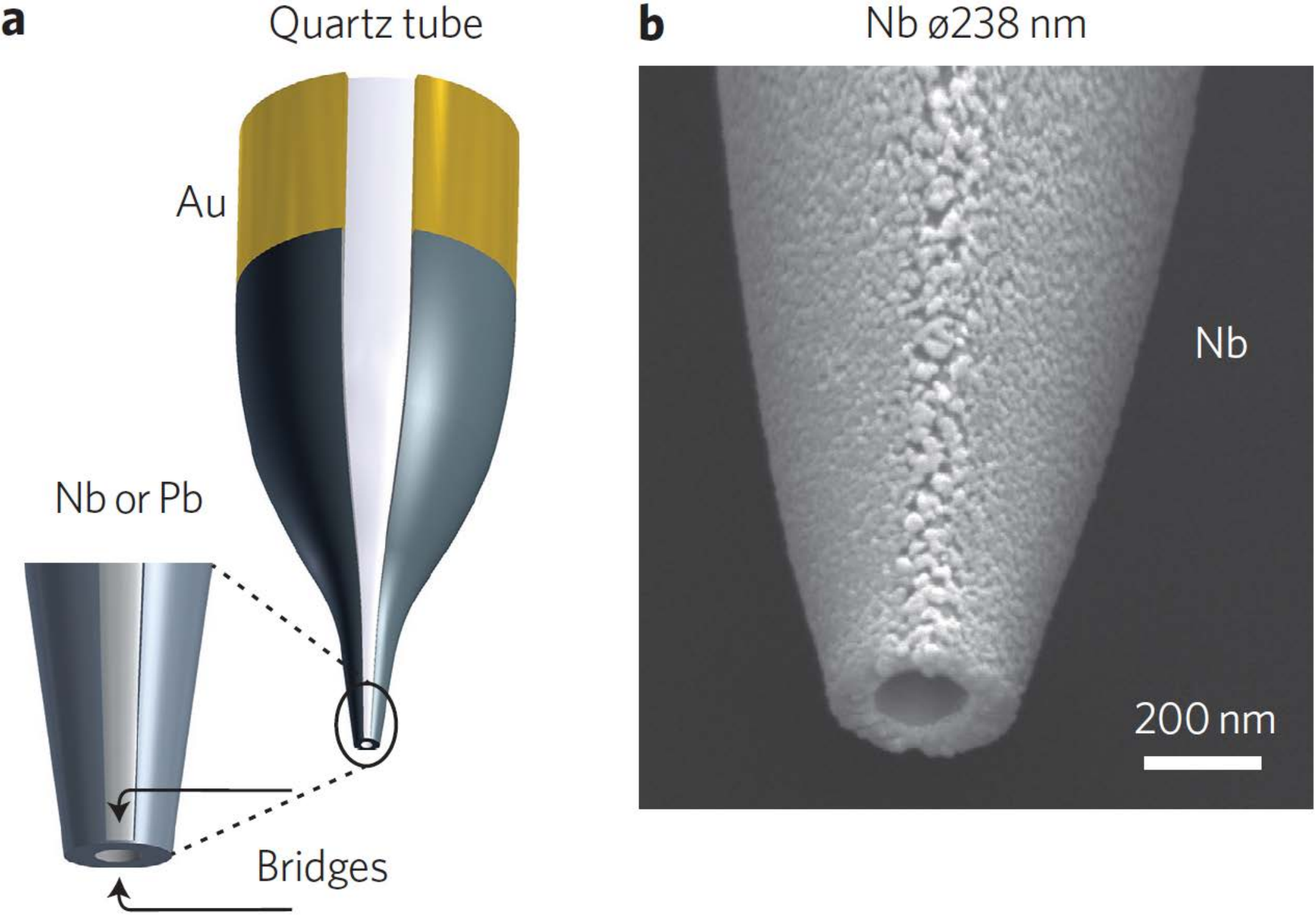}
%
\caption{SQUID-on-tip (SOT):
(a) schematic of a sharp quartz pipette with superconducting leads, connecting to the SOT at the bottom end; inset shows magnified view.
(b) SEM image of a Nb SOT having a diameter of 238\,nm.
Reprinted by permission from Macmillan Publishers Ltd: Nature Nanotechnology \cite{Vasyukov13}, copyright (2013).}
\label{Fig:20}
%
\end{figure}

An important breakthrough in the field of nanoSQUIDs applied to SSM was achieved recently with the implementation of the SQUID-on-tip (SOT) by the Zeldov group at the Weizman Institute of Science \cite{Finkler10,Vasyukov13}.
This device is based on the deposition of a nanoSQUID directly on the apex of a sharp quartz pipette [Fig.\ref{Fig:20}].
The fact that the nanoSQUID is located on a sharp tip reduces the possible minimum probe-to-sample distances to below 100\,nm, boosting enormously the spatial resolution of the microscope.
Al, Nb and Pb nanoSQUIDs based on Dayem bridges are shadow-evaporated in a three-angle process, without requiring any lithography or milling steps.
For this purpose, a quartz pipette is first pulled to form a sharp hollow tip with $40-300\,$ nm inner diameter.
By means of a laser diode parallel to the tip, the latter is aligned pointing down towards the source which defines the $0^\circ$ position.
Then a thin layer ($< 10\,$nm) of superconducting material is deposited, followed by two thicker leads ($> 25\,$nm) deposited at $\pm 100^\circ$.
The resulting weak links formed at the tip apex between these two leads constitute two Dayem bridges.
Special care must be taken for fabricating the Nb and Pb sensors.
The former require the previous deposition of a thin AlO$_x$ buffer layer to prevent contamination from the quartz tip.
A dedicated ultra-high vacuum e-beam evaporation system was used for depositing Nb from a point source.
On the other hand, the so far most sensitive Pb sensors require the use of a He cooling system for the tips during deposition to prevent the formation of islands due to the large surface mobility of these atoms at higher temperatures.
This procedure lead to the smallest nanoSQUIDs fabricated so far, with effective nanoloop diameters down to 50\,nm.
The resulting inductance of the loop reaches values below 10\,pH, dominated by the kinetic inductance of the thin superconducting layer.
Although these nanoSQUIDs exhibit hysteretic IVCs, operation with voltage bias and reading out the resulting current signal with an SSA enables the detection of the intrinsic flux noise of the devices.
The SOTs can be operated in large magnetic fields up to $\sim 1\,$T (limited by the upper critical fields of the superconducting materials).
So far, flux biasing to maintain the optimum working point during continuous external field sweep is not possible.
By adjusting the external magnetic field to values which yield large transfer functions, these devices exhibit extraordinary low flux noise levels down to $50\,{\rm n}\Phi_0/\sqrt{\rm Hz}$ for the Pb SOTs \cite{Vasyukov13}.
The latter varies, depending on the biasing external magnetic field.
For a magnetic dipole located at the center of the loop with orientation perpendicular to the loop plane (assuming an infinitely narrow width of the loop, i.e.~the approximation used by Ketchen {\it et al.} \cite{Ketchen89}), this translates into a spin sensitivity of $0.38\,\mu_{\rm B}/\sqrt{\rm Hz}$, i.e.~the best spin sensitivity reported so far for a nanoSQUID.

A device able of distinguishing in-plane and out-of-plane magnetic signals was also reported \cite{Anahory14}.
This is achieved by using a pipette with $\theta$-shaped cross section to form a three JJ SQUID (3JSOT).
This tip is later milled by FIB leading to a V shaped apex with two oblique nanoloops connected in parallel.
By measuring the dependence of the maximum critical current on the externally applied in-plane and out-of-plane magnetic fields $I_c(H_{||},H_\perp$), it is possible to determine all the geometrical and electric parameters of the device.
Field components can be decoupled by biasing the 3JSOT at specific fields ($H_{||},H_\perp$) in which $I_c$ depends strongly on one of the two orthogonal components of the magnetic field while being insensitive to the other.
As a drawback, this device is not able of distinguishing both in-plane and out-of-plane components of the magnetic flux simultaneously, but only when operated at different flux biasing points.

For SSM, a system operating in a $^3$He system with 300\,mK base temperature has been developed, with the SOT glued on a quartz tuning fork, to operate the system also in an magnetic force microscopy mode.
This allows scanning (using piezo-scanners) at extremely small tip-to-sample distances of only a few nm \cite{Finkler12}.
A spatial resolution below 120\,nm was demonstrated by imaging vortices in Nb thin films with a 117\,nm-diameter Pb SOT \cite{Vasyukov13}.

The SOT-SSM system has been successfully applied to the study of vortex trajectories in superconducting thin films, allowing the investigation of the influence of the pining force landscape \cite{Embon15}.
More recently, this tool was used to observe nanoscopic magnetic structures such as ferromagnetic metallic nanoislands at the LaMnO$_3$/SrTiO$_3$ interface \cite{Anahory15} or magnetic nanodomains in magnetic topological insulators \cite{Lachman15}.

\section{Summary and outlook}
\label{sec:Summary}

Significant progress in thin film fabrication and pattering technologies has enabled the development of strongly miniaturized dc SQUIDs with loop sizes on the micrometer scale (microSQUIDs) or even with sub-micrometer dimensions (nanoSQUIDs), or SQUIDs coupled to miniaturized pickup loops.
Such devices are based on a variety of Josephson junctions, intersecting the SQUID loop, many of them also on the sub-micrometer scale.
As a key advantage of such strongly miniaturized SQUID structures, they can offer significantly reduced flux noise,
down to the level of a few tens of n$\Phi_0/\sqrt{\rm Hz}$, corresponding to spin sensitivities around $1 \mu_{\rm B}/\sqrt{\rm Hz}$ and improved spatial resolution for scanning SQUID microscopy.
Hence, strongly miniaturized SQUIDs are very promising detectors for investigating tiny and strongly localized magnetic signals produced, e.g., by magnetic nanoparticles or for high-resolution scanning SQUID microscopy.
Very recent advances, including the demonstration of single spin sensitivity and a breakthrough in spatial resolution of scanning SQUID microscopy open up promising perspectives for applications in nanoscale magnetism of condensed matter systems.


\acknowledgments

We gratefully acknowledge valuable contributions by the nanoSQUID team at T\"{u}bingen, M.~Kemmler, J.~Nagel, R.~W\"{o}lbing, T.~Schwarz, B.~M\"{u}ller, S.~Hess and R.~Kleiner, and by our collaborators O.~Kieler and A.~Zorin {\it et al.} at PTB Braunschweig, T.~Schurig {\it et al.} at PTB Berlin, B.~B\"{u}chner {\it et al.} at IFW Dresden, M.~Poggio {\it et al.} at Univ.~Basel, D.~Grundler {\it et al.} at TU Munich and EPFL Lausanne, A.~Fontcuberta i Morral at EPFL Lausanne and J.~Ses\'{e} at INA Zaragoza.
This work was funded by the Alexander von Humboldt Foundation, the Nachwuchswissenschaftlerprogramm of the Univ.~T\"{u}bingen, the Deutsche Forschungsgemeinschaft, via projects KO 1303/13-1 and SFB/TRR21, and by the EU-FP6-COST Action MP1201.
  

\bibliography{nanoSQUID_WdG}
\end{document}